\documentclass[
nofootinbib,
superscriptaddress,
preprint,
amsmath,amssymb,
aps,
prd,
]{revtex4-2}

\usepackage{graphicx}
\usepackage{dcolumn}
\usepackage{bm}
\usepackage{hyperref}
\usepackage{slashed}
\usepackage{microtype}
\usepackage{bbm} 
\newcommand{\bx}{\mathbf{x}}
\newcommand{\by}{\mathbf{y}}
\newcommand{\bp}{\mathbf{p}}
\newcommand{\bq}{\mathbf{q}}
\newcommand{\bk}{\mathbf{k}}
\newcommand{\bn}{\mathbf{n}}
\providecommand{\abs}[1]{\lvert#1\rvert}
\allowdisplaybreaks

\begin{document}
\preprint{HUPD-2102}
\title{
  Spacetime evolution of lepton number densities and wave packet-like effects for neutrino flavor and chiral oscillations in quantum field theory
}

\author{Apriadi Salim Adam}
\email{apriadi.salim.adam@brin.go.id}%
\affiliation{%
Research Center for Quantum Physics, National Research and Innovation Agency (BRIN),  South Tangerang 15314, Indonesia
}%
\author{Nicholas J. Benoit}%
\email{njbenoit@hiroshima-u.ac.jp}
\affiliation{%
Physics Program, Graduate School of Advanced Science and Engineering, 
Hiroshima University, Higashi-Hiroshima 739-8526, Japan
}%
\author{Yuta Kawamura}%
\email{kawamura1994phy@gmail.com}
\affiliation{%
Kitakami city,  Iwate,  Japan
}%
\author{Yamato~Matsuo}%
\email{yamatsuos0@gmail.com}
\affiliation{%
Akashi city,  Hyogo, Japan
}%
\author{Takuya Morozumi}%
\email{morozumi@hiroshima-u.ac.jp}%
\affiliation{%
Physics Program, Graduate School of Advanced Science and Engineering,
Hiroshima University, Higashi-Hiroshima 739-8526, Japan
}%
\affiliation{%
Core of Research for the Energetic Universe, Hiroshima University, \\
Higashi-Hiroshima 739-8526, Japan
}%
\author{Yusuke Shimizu}%
\email{shimizu.yusuke@kaishi-pu.ac.jp}%
\affiliation{%
Department of Information, Kaishi Professional University, 
Niigata 950-0916, Japan
}%
\author{Naoya Toyota}%
\email{zhizailitian7@gmail.com}%
\affiliation{%
Osaka,  Japan
}%


\begin{abstract} 
We present a formulation of lepton family numbers,  based on quantum field theory, for neutrino oscillation phenomenology that can be applied to nonrelativistic and relativistic energies for neutrinos.
It is formulated for both types of neutrinos, Dirac and Majorana.
The formulation is constructed as the time evolution of a lepton family number density operator. 
Then, the time evolution of the lepton family number density operator becomes dependent on the mass and new features appear.
The expectation value of the density operator is evaluated for the initial state with a Gaussian distribution for the momentum amplitude.
This enables us to study wave packet-like decoherence effects.
We show in the nonrelativistic regime, the type of neutrino mass are distinguishable even under the presence of wave packet-like decoherence effects.
\end{abstract}

\maketitle

\section{\label{sec:Intro} Introduction}
Neutrinos were originally formulated, within the weak interactions of the Standard Model (SM), to be massless fermions.
However, hints of massive neutrinos appeared in the second half of the 20th century with the solar neutrino problem\cite{Bahcall:2002ng}.
Eventually, updated solar models and experiments from the Kamiokande laboratory and the Sudbury Neutrino Observatory (SNO) proved the disappearance is caused by flavor oscillations\cite{Abe:2010hy,Aharmim:2009gd}.
For neutrinos the existence of flavor oscillations mean there is a mixing between mass and flavor eigenstates.
The mixing between mass and flavor is governed by the unitary PMNS matrix, which has six (four) free parameters in the $3\times3$ instance\cite{Pontecorvo:1957qd,Maki:1962mu}.
Thus, the mixing caused by flavor oscillations is direct evidence that neutrinos are massive and, consequently, require physics beyond the SM.

In this decade, the six parameters related to the PMNS matrix; the oscillation angles $\theta_{12},\theta_{23}, \text{and }\theta_{13}$, the mass squared differences $\Delta m_{21}^2\text{ and }\Delta m^2_{31}$, and the Dirac CP violating phase $\delta_{cp}$, are expected to be measured within a few percentages by neutrino oscillation experiments\cite{Abi:2020evt,Abe:2015zbg,Abe:2016ero,Baussan:2013zcy,Kumar:2017sdq,An:2015jdp}.
In addition, the absolute mass of the lightest neutrino, $m_1$ or $m_3$, is expected to be directly limited down to the sub-eV range by the Karlsruhe Tritium Neutrino Experiment (KATRIN)\cite{Aker:2019qfn}.
However, even with these precision measurements, questions remain in neutrino phenomenology.
Specifically, the question of what type of mass neutrinos possess, Dirac or Majorana remains.
This question could be answered by future neutrinoless double-$\beta$ $(0\nu\beta\beta)$ decay experiments\cite{Dolinski:2019nrj,Furry:1939qr,Bahcall:1978jn,Schechter:1980gk,Xing:2013ty}.
Arguably, additional approaches to answer those questions would be ideal.

For this work, we present a unified description of neutrino phenomenology that leads to an additional approach for investigations of the previously mentioned questions.
We define unified to mean a description that includes neutrino flavor oscillations and, particle-anti-particle or chiral oscillations.
In addition, our description is different from the usual neutrino oscillation theory that assumes neutrinos are relativistic.
The relativistic assumption is often taken, because cosmological limits on the neutrinos absolute masses place them in the sub-eV range\cite{Aghanim:2018eyx}; whereas, experiments are performed on neutrinos with energies in the keV, MeV, GeV, and recently the PeV ranges\cite{Formaggio:2013kya}.
Nonrelativistic neutrinos are predicted to exist in nature as the cosmic neutrino background (C$\nu$B).
Experiments to detect the C$\nu$B are under preparation (see for example, \cite{Betti:2019ouf}), and we plan future studies of the C$\nu$B as an application of our description.

For studies of the C$\nu$B we need to build a framework that uses a momentum distribution \cite{Dolgov:2002wy}.
Since the state of the C$\nu$B can be described by a mixed state at their decoupling time, their momentum distribution should be given by a density matrix.
For oscillation experiments, if the neutrino is produced with a localized space, it should be described by a pure state with a momentum distribution.
Descriptions of that type have been done using wave packet formulations \cite{Giunti:1991ca,Akhmedov:2019iyt,Akhmedov:2017xxm,Akhmedov:2009rb,Nussinov:1976uw}.
To incorporate a momentum distribution in our framework, we consider densities of lepton family numbers that are localized in space as pure states.
We have previously discussed details of the formulation for the Majorana mass case using plane waves \cite{Adam:2021qiq,Adam:2021vbl,Benoit:2022dsc}.

In future work we can extend this study to the mixed state required by C$\nu$B, and further consider the time evolution of lepton numbers under the expansion of the universe.
Then, we may be able to clarify when the C$\nu$B transits from the relativistic regime to the nonrelativistic regime.
One may also find how the coherence for a given momentum distribution will continue or will be lost as the C$\nu$B is redshifted due to the expansion.
Therefore, our present work can be useful to clarify those natures of the C$\nu$B. 

\section{Brief outline of related studies}
Consider the decay of a positively charged pion at rest to an anti-muon and neutrino,
\begin{equation}
    \pi^+ \rightarrow \mu^+ + \nu_\alpha.
    \label{eq:chargedpiondecay}
\end{equation}
The Standard Model defines the neutrino flavor, $\alpha$, to be a muon-neutrino $\nu_\mu$.
This definition is from connecting the neutrino and charged lepton with the electroweak doublets, i.e., $(\nu_e,e)$, $(\nu_\mu,\mu)$, and $(\nu_\tau,\tau)$.
Then, theory introduced lepton family numbers as a way to express those connections; $L_e$, $L_\mu$, and $L_\tau$ \cite{Konopinski:1953gq,Pontecorvo:1959sn}.
A consequence is the Standard Model conserves lepton family numbers in all processes.

Yet, neutrino masses imply lepton family numbers are not necessarily conserved \cite{Bilenky:2001yh}.
This is evident in neutrino oscillation experiments.
Consider the example of Eq.(\ref{eq:nuoscexperiments}), which uses the produced anti-muon to identify a muon neutrino at the production point.
Then after the neutrino propagates over a macroscopic distance, either a muon neutrino or electron neutrino is identified at a detector.
\begin{equation}
    \pi^+ \rightarrow \mu^+ + \nu_\mu \xrightarrow{\text{propagation}}
    \begin{cases}
        \nu_\mu + X \rightarrow \mu^- + Y & \text{disappearance,}\\
        \nu_e + X \rightarrow e^- + Y & \text{appearance.}
    \end{cases}
  \label{eq:nuoscexperiments}
\end{equation}
If we compare the family of the charged leptons from production to the appearance detection, a change of $L_\mu$ to $L_e$ occurs.
Models explain this change by treating the interacting and propagating neutrinos differently.

In the quantum mechanics model, the interacting neutrinos are from a flavor basis and the propagating neutrinos are from a mass basis \cite{Kayser:1981ye,Giunti:2000kw}.
Those bases are related to each other by a unitary transformation with a unitary matrix called the Pontecorvo-Maki-Nakagawa-Sakata (PMNS) matrix, $U^\ast_{\alpha i}$.
Then, a neutrino flavor state $\vert \nu_\alpha \rangle$ is written as a coherent superposition of massive states weighted by the PMNS matrix,
\begin{equation}
    \vert \nu_\alpha \rangle = \sum^3_{i=1} U^\ast_{\alpha i} \vert \nu_i \rangle.
    \label{eq:pontecorvostates}
\end{equation}
Each forms an orthogonal basis $ \langle \nu_\alpha | \nu_\beta \rangle =\delta_{\alpha \beta}$  and $ \langle \nu_i | \nu_j \rangle=\delta_{i, j}$.
Sometimes these flavor states are called Pontecorvo states in the literature.
However, it was shown that the states of Eq.(\ref{eq:pontecorvostates}) can not be applied to quantum field theory \cite{Giunti:1991cb,Blasone:2019rxl,Blasone:2006jx}.
Consequently, effort has been made to understand the nature of flavor in quantum field theory.
We introduce some quantum field theory models that are relevant to our work.

A common quantum field theory model treats the produced or detected neutrino flavor state is different from the Pontecorvo states in Eq.(\ref{eq:pontecorvostates}).
The difference appears in the amplitudes of the superposition to each mass eigenstate.
They are not given by the PMNS matrix elements alone.
They depend on the additional factor coming from either the production or detection amplitude of each mass eigenstate.
Then, that model derives the neutrino flavor states based on the production or detection processes \cite{Giunti:2002xg,Giunti:2004zf}.
As a result, the neutrino flavor states are process dependent and modify the amplitudes of neutrino flavor oscillation probabilities.
In the appearance experiment of Eq.(\ref{eq:nuoscexperiments}), it also results in the detected neutrino flavor state and the produced neutrino flavor state being non-orthogonal to each other even before oscillations start.
Nevertheless, if the neutrinos are treated as ultra-relativistic particles, the modifications are lost.

In a different model, wave packets for the external particles are introduced, and the neutrinos appear as intermediate states between production and detection \cite{Giunti:1993se,Beuthe:2001rc,Asahara:2004mh,Akhmedov:2010ms,Wu:2010yr}.
For this model, no neutrino flavor states are considered or calculated.
Only the source current of the neutrino and the detector current appear as physical particles.
If the method is applied to Majorana neutrinos, two processes are required. 
The first process is to detect the chirality and lepton number conserving propagation.
The second process is for chirality and lepton number violating propagation of the intermediate neutrinos.  
In our framework,  both of the chirality conserving and violating effects are included in the lepton family numbers.

Lastly, another model considers the neutrino flavor states to be from a physical Fock space.
The resulting neutrino flavor states are related to massive neutrino states by a Bogoliubov transformation, which leads to inequivalent vacua \cite{Blasone:2002jv,Blasone:2001du,Blasone:2001qa,Fujii:1998xa,Fujii:2001zv}.
This model introduces non-zero Dirac mass to each flavor neutrino field \cite{Giunti:2003dg,Tureanu:2020odo,Blasone:2020wer}.
A flavor charge is defined through vector current of the flavor neutrino.
In our framework, the flavor charge for active neutrinos is defined based on the weak eigenstates that form SU(2) doublets with the mass eigenstates of the charged leptons.
Therefore, the charge is defined through the left-handed current of the weak eigenstates.
We call this charge lepton family number.
Then in the formula for the lepton family number, only the neutrino masses for the mass eigenstates appear and additional mass parameters for the flavor neutrinos are not introduced.

\section{Lepton Number Density}
We take an approach to define lepton family numbers for neutrinos based on the isospin doublet used for charged current weak interactions.
Meaning, in the diagonal basis for charged lepton mass, the lepton family number is assigned to each lepton doublet $\Psi_{L \alpha}=\begin{pmatrix} \nu_{L \alpha} &  e_{L \alpha} \end{pmatrix}^T $ as well as the right-handed charged lepton $e_{R \alpha}$.  
Definitely, under a $U_{\alpha}$(1) transformation of the lepton family number $L_\alpha$, they transform as,
\begin{align}
  \Psi^\prime_{L \alpha}=\begin{pmatrix} \nu^\prime_{L \alpha} \\ e^\prime_{L \alpha}  \end{pmatrix}  = e^{i \theta_\alpha}  \Psi_{L \alpha}=e^{i \theta_\alpha} \begin{pmatrix} \nu_{L \alpha} \\ e_{L \alpha}  \end{pmatrix} , && e^\prime_{R \alpha}= e^{i \theta \alpha} e_{R \alpha}.
\end{align}
Where the subscript $L$ and $R$ denote the left-handed projection operator $P_L=\frac{1-\gamma^5}{2}$
and $P_R=\frac{1+\gamma^5}{2}$, respectively.
The associated Noether current is defined by
\begin{equation}
  l^{Noether}_{\mu \alpha}=: \overline{e_\alpha} \gamma_\mu e_{\alpha} :+ :\overline{\nu_{\alpha}} \gamma_\mu P_L \nu_{\alpha}:,
\end{equation}
where the colon : implies  normal ordering.
Note, the lepton family current consists of a vector current of the charged lepton and a left-handed current of the neutrino.
We describe the effects of mass of neutrinos on lepton number in this particular weak basis where the charged lepton mass matrix is real and diagonal.
Through the detection of a charged lepton flavor, the state of the neutrino is projected to the flavor associated to the charged lepton.
In the process, the lepton family number is assumed to be conserved.
For the example,  $\pi^+ \rightarrow \mu^+ +\nu $, the produced neutrino is assumed to have a muon number $+1$.
The projected state has a left-handed chirality with a definite lepton family number.
We will see that the neutrino field with definite chirality can be expanded by the plane wave solution and spinors with helicity $-\frac{1}{2}$.
The effect of the non-zero mass of neutrino is encoded in non-trivial time dependence of creation and annihilation operators which we will evaluate below.

\subsection{Majorana neutrino formulation}\label{sec:majoranaFormulation}
In the flavor basis where the charged lepton mass matrix is real and diagonal, the free part of the Lagrangian for Majorana neutrinos is,
\begin{equation}
  \mathcal{L}^M = \overline{\nu_{L\alpha}} i \gamma^\mu \partial_\mu \nu_{L\alpha}-\frac{1}{2}\left(\overline{ (\nu_{L \alpha})^C} m_{\alpha\beta}\nu_{L\beta} + \overline{\nu_{L \alpha}}(m^\ast_{\alpha\beta}) (\nu_{L \beta})^C \right).
  \label{eq:LagMajoranamulti1}
\end{equation}
The equation of motion for the Majorana neutrino in the flavor basis is written as;
\begin{equation}
  \begin{split}
    i \slashed{\partial}\nu_{L \alpha} &= m^\ast_{\alpha \beta} (\nu_{L \beta})^{C}, \\
    i \slashed{\partial} (\nu_{L \alpha})^C &= m_{\alpha \beta} (\nu_{L \beta}), 
  \end{split}
  \label{eq:physicaleqofmot}
\end{equation}
where $m_{\alpha \beta}$ is a Majorana mass matrix.
The Greek subscripts represent the flavors $e$, $\mu$, and $\tau$.
In the flavor basis, the Majorana mass matrix is complex and symmetric.
In addition, we have used the notation $(\nu_{L\alpha})^{C}$ for charge conjugation.
The solution of Eq.(\ref{eq:physicaleqofmot}) can be written with the form of the expansion with the massless spinors plus a zero-mode contribution.
The details on how we quantize this situation is written in appendix \ref{sec:zeromodes}.
The remainder of the main text will focus on the massless spinor, i.e., the non-zero-mode contributions,
\begin{equation}
  \nu_{L \alpha}(t, \bx)= 
  \int^\prime \frac{d^3\bp}{(2\pi)^32|\bp|}\left(a_{\alpha}^{}(\bp, t )u_L(\bp)e^{i\bp\cdot\bx}+b^\dagger_{\alpha}(\bp, t)v_L(\bp)e^{-i\bp\cdot\bx}\right),
  \label{eq:initialc}
\end{equation}
where $\int^\prime$ implies the exclusion of zero momentum mode from the integration. 
The massless spinors $u_L(\bp)$ and $v_L(\bp)$ are given by \cite{Adam:2021qiq},
\begin{gather}
   u_L(\bp)=-v_L(\bp)=\sqrt{\abs{2\bp}}\begin{pmatrix} 0 \\ \phi_-(\bn) \end{pmatrix},
   \label{eq:spinormathn} \\
   \bn \cdot \boldsymbol{\sigma} \phi_\pm(\bn)=\pm \phi_\pm(\bn), \quad \bn = \frac{\bp}{\abs{\bp}},
\label{eq:pointingvectorn}
\end{gather}
where the two component spinors $\phi_\pm(\pm\bn)$ are written by the polar angle $\theta$ and the azimuthal angle $\phi$ specifying the direction of the momentum 
$\bp$,
\begin{align}
  \phi_+(\bn) & = \begin{pmatrix} e^{-i \frac{\phi}{2}} \cos\frac{\theta}{2} \\ e^{i \frac{\phi}{2}} \sin\frac{\theta}{2} \end{pmatrix}, &
  \phi_-(\bn) & = \begin{pmatrix} - e^{-i \frac{\phi}{2}} \sin\frac{\theta}{2} \\ e^{i \frac{\phi}{2}} \cos\frac{\theta}{2} \end{pmatrix},
\\
  \phi_+(-\bn) & = i \phi_-(\bn), &
  \phi_-(-\bn) & = i \phi_+(\bn).
\end{align}
We then formulate a Heisenberg operator for $L^M_{\alpha}$  where $\alpha=(e,\mu,\tau)$, from the lepton number density  $l^M_{\alpha}(t,\bx)$ by integrating over space, 
\begin{gather}
  L^M_{\alpha}(t)=\int d^3x \, l^M_{\alpha}(t,\bx)=\int^{\prime} \frac{d^3 k}{(2\pi)^3 2\abs{\bk}}
  \left[a^\dagger_\alpha(\bk,t)a_\alpha(\bk,t)-b^\dagger_\alpha(\bk,t)b_\alpha(\bk,t)\right],
  \label{Eq:LeptonNumber}
\\
  l^M_{\alpha}(t,\bx) = \, :\overline{\nu_{L\alpha}}(t,\bx)\gamma^0 \nu_{L\alpha}(t,\bx):\,,
  \label{Eq:LeptonNumberDensity}
\end{gather}
where the colon : denotes normal ordering.
In the absence of mass matrix, the time evolution of creation and annihilation operators are
\begin{align}
  a_\alpha(\bp, t)=a_\alpha(\bp, t_0) e^{-i|\bp|(t-t_0)}, &&
  b^\dagger_\alpha(\bp, t)=b^\dagger_\alpha(\bp,t_0) e^{i|\bp|(t-t_0)}.
\end{align}
In this case, all three lepton family number operators are separately conserved,
\begin{equation}
  L^M_\alpha(t)=L^M_\alpha(t_0)
\end{equation}
However as can be seen from  Eq.(\ref{eq:physicaleqofmot}), their time dependence is governed by non-diagonal mass matrix.
To obtain the time dependence of the lepton number operator in Eq.(\ref{Eq:LeptonNumber}), we need to know the time dependence of the flavor operators $ a_\alpha(\bk,t)$ and $b_\alpha(\bk,t)$.
In order to evolve the operators to time $t$ from the initial time $t_0$, one needs to express them at $t_0$ as a superposition of operators with definite masses.
This is because the operators with definite masses $m_i$ evolve like $e^{\pm i E_i t}$, with $E_i=\sqrt{|\bp|^2+m_i^2}$.
Then the time dependence of the flavor operator can be derived.

We begin by diagonalizing the mass matrix in Eq.(\ref{eq:physicaleqofmot}) with a unitary matrix
$V$ as,
\begin{gather} \label{Eq:majoranadiagonal}
  m_{i}\delta_{ij}=\left(V^T\right)_{i\alpha}m_{\alpha\beta}V_{\beta j},\\
  \nu_{L\alpha}=V_{\alpha i}\nu_{L i},
\end{gather}
where $\nu_{L i}$ denotes an operator of the mass basis.
$\nu_{L i}$ and its charge conjugation form the operator for the Majorana field as,
\begin{equation}
  \begin{split}
    \psi_{M i}({\bf x}, t)&= \nu_{L i}({\bf x}, t)+ ( \nu_{L i}({\bf x}, t))^c \\
    &= V_{\alpha i}^\ast \nu_{\alpha L} ({\bf x}, t)+V_{\alpha i} (\nu_{\alpha L}({\bf x}, t) )^c.
  \end{split}
  \label{eq:MajoranaVSmassless}
\end{equation}
Since Eq.(\ref{eq:MajoranaVSmassless}) satisfies the Majorana condition, it can be expanded as
\begin{equation}
  \psi_{M i}({\bf x}, t) = \int^\prime \frac{d^3\bp}{(2\pi)^32E_i(\bp)}\sum_{\lambda=\pm}\Bigl(a^{}_{M i}(\bp,\lambda)u_i(\bp,\lambda)e^{-i (E_i t - \bp\cdot\bx) }+a^\dagger_{M i}(\bp,\lambda) v_i(\bp,\lambda)e^{i (E_i t -\bp \cdot \bx) }\Bigr).
  \label{eq:Majoranaexp}
\end{equation}
where $E_i=\sqrt{|\bp|^2+m_i^2}$ denotes the energy of the mass eigenstate and $u_i(\bp,\lambda) $ and $v_i(\bp,\lambda)$ denote the massive Dirac spinors with the definite helicity $\lambda$ and mass $m_i$.
For the details of the definitions, see appendix A in Ref.\cite{Adam:2021qiq}.
The operators $a^{}_{Mi}(\bp,\lambda)$ of Eq.(\ref{eq:Majoranaexp}) are distinct from the operators with the massless spinors in Eq.(\ref{eq:initialc}).
We will call $a^{}_{Mi}(\bp,\lambda)$ the Majorana operators, and they obey the anti-commutation relation,
\begin{equation}
    \{a^{}_{Mi}(\bp,\lambda),a^\dagger_{Mj}(\bq,\lambda^\prime)\} = 2E_i (\bp)(2\pi)^3\delta^{(3)}(\bp-\bq)\delta_{ij}\delta_{\lambda\lambda^\prime}\,,
\end{equation}
with all others being zero.

Using Eq.(\ref{eq:MajoranaVSmassless}) with the expansions Eq.(\ref{eq:Majoranaexp}) and Eq.(\ref{eq:initialc}), one can derive the relation between the  operators $a_\alpha(\bp, t), b_\alpha(\bp, t)$ and the operators for massive fields $a_{M i} (\bp, \lambda)$ for arbitrary time $t$.
\begin{gather}\label{eq:flavorOpa}
  a^{}_{\alpha}(\pm \bp, t)=V_{\alpha i}\frac{\sqrt{2\abs{\bp}(E_i(\bp)+\abs{\bp})}}{2E_i(\bp)}\left(a_{Mi}(\pm \bp,-) e^{-i E_i(\bp) t}\pm \frac{im_i}{E_i(\bp)+\abs{\bp}}a^\dagger_{Mi}(\mp \bp,-)  e^{i E_i(\bp) t}\right),
\\ \label{eq:flavorOpb}
  b^{}_{\alpha}(\pm \bp, t)=V_{\alpha i}\frac{\sqrt{2\abs{\bp}(E_i(\bp)+\abs{\bp})}}{2E_i(\bp)}\left(a_{Mi}(\pm \bp,+)  e^{-i E_i(\bp) t} \pm \frac{im_i}{E_i(\bp)+\abs{\bp}}a^\dagger_{Mi}(\mp \bp,+) e^{i E_i(\bp) t}\right),
\end{gather}
where $+\bp$ denotes the momentum directed toward the positive y-axis hemisphere $(0 \le \phi < \pi) $ and $-\bp$ denotes the momentum directed toward the negative y-axis hemisphere $(\pi \le \phi < 2\pi) $. 
Notice, the relations are a non-trivial mixing of the Majorana annihilation and creation operators forming $a_\alpha(\pm \bp, t)$ and $b_\alpha(\pm \bp,t )$.
Furthermore, we can identify the relations as a Bogoliubov transformation that occurs between the operators \cite{Morozumi:2022mqh}.
The operators $a^{}_{\alpha}(\pm \bp, t)$ and $b^{}_{\alpha}(\pm\bp, t)$, in Eq.(\ref{eq:flavorOpa}) and Eq.(\ref{eq:flavorOpb}), satisfy the anti-commutation relations,
\begin{equation}
  \{a^{}_{\alpha}(\pm \bp, t),a^\dagger_{\beta}(\pm \bq,t )\} =
  \{b^{}_{\alpha}(\pm \bp, t),b^\dagger_{\beta}(\pm \bq,t )\} 
  = 2\abs{\bp}(2\pi)^3\delta^{(3)}(\bp-\bq)\delta_{\alpha\beta}\,,
  \label{eq:masslessAnticommutation}
\end{equation}
with all other relations being zero. 
The inverse relations at $t=t_0$ are also obtained as,
\begin{gather}
  a_{Mi}(\pm \bp, -) = e^{i E_i(\bp) t_0}\sqrt{\frac{E_i(\bp)+\abs{\bp}}{2\abs{\bp}}} \left( V^\ast_{\beta i}a^{}_{\beta}(\pm \bp, t_0)\mp V_{\beta i} \frac{i m_i}{E_i(\bp)+\abs{\bp}}  a^{\dagger}_{\beta}(\mp \bp, t_0)\right), 
  \label{eq:inverse1}
\\
  a^{\dagger}_{Mi}(\pm \bp, -) = e^{-i E_i(\bp) t_0}\sqrt{\frac{E_i(\bp)+\abs{\bp}}{2\abs{\bp}}} \left( V_{\beta i}a^{\dagger}_{\beta}(\pm \bp, t_0)\pm V^\ast_{\beta i} \frac{i m_i}{E_i(\bp)+\abs{\bp}}  a^{}_{\beta}(\mp \bp, t_0)\right),
  \label{eq:inverse2}
\\
  a_{Mi}(\pm \bp, +) = e^{i E_i(\bp) t_0}\sqrt{\frac{E_i(\bp)+\abs{\bp}}{2\abs{\bp}}} \left( V_{\beta i} {b}_{\beta}(\pm \bp, t_0)\mp V^\ast_{\beta i} \frac{i m_i}{E_i(\bp)+\abs{\bp}}  b^{\dagger}_{\beta}(\mp \bp, t_0)\right),
  \label{eq:inverse3} 
\\
  a^{\dagger}_{Mi}(\pm \bp, +) = e^{-i E_i(\bp) t_0}\sqrt{\frac{E_i(\bp)+\abs{\bp}}{2\abs{\bp}}} \left( V^\ast_{\beta i}b ^{\dagger}_{\beta}(\pm \bp, t_0)\pm V_{\beta i} \frac{i m_i}{E_i(\bp)+\abs{\bp}}  b^{}_{\beta}(\mp \bp, t_0)\right).
  \label{eq:inverse4}
\end{gather}
By substituting the inverse relations Eqs.(\ref{eq:inverse1}-\ref{eq:inverse4}) into Eq.(\ref{eq:flavorOpa}) and Eq.(\ref{eq:flavorOpb}), we obtain relations between the arbitrary time operators $(a_\alpha(\pm \bp, t), b_\alpha(\pm \bp ,t))$ and the initial time operators $(a_\alpha(\pm \bp, t_0), b_\alpha(\pm \bp ,t_0))$.
\begin{gather}
  \begin{split} \label{eq:timeDependOpa}
    a^{}_{\alpha}(\pm \bp, t)=\sum_{\beta=e}^\tau\sum_{j}\biggl(V^{}_{\alpha j}V^\ast_{\beta j}&\left[\cos[{E_j(\bp)(t-t_0)}]-i\frac{\abs{\bp}}{E_j(\bp)}\sin[{E_j(\bp)(t-t_0)}]\right]a^{}_\beta(\pm \bp, t_0)\biggr.
  \\
    &\biggl.\mp V^{}_{\alpha j}V^{}_{\beta j}\frac{m_j}{E_j(\bp)}\sin[{E_j(\bp)(t-t_0)}]a^\dagger_\beta(\mp \bp, t_0)\biggr),
\raisetag{40pt} 
  \end{split}
\\
  \begin{split} \label{eq:timeDependOpb}
    b^{}_{\alpha}(\pm \bp, t)=\sum_{\gamma=e}^\tau\sum_{j}\biggl(V^\ast_{\alpha j}V^{}_{\gamma j}&\left[\cos[{E_j(\bp)(t-t_0)}]-i\frac{\abs{\bp}}{E_j(\bp)}\sin[{E_j(\bp)(t-t_0)}]\right]b^{}_\gamma(\pm \bp, t_0)\biggr.
  \\
    &\biggl.\mp V^\ast_{\alpha j}V^\ast_{\gamma j}\frac{m_j}{E_j(\bp)}\sin[{E_j(\bp)(t-t_0)}]b^\dagger_\gamma(\mp \bp, t_0)\biggr).
  \end{split}
\end{gather}
The equations (\ref{eq:timeDependOpa}) and (\ref{eq:timeDependOpb}) tell us the relations between the operators $(a_\alpha(\pm \bp, t),  b_\alpha(\pm \bp, t))$ and $(a_\alpha(\pm \bp, t_0),  b_\alpha(\pm \bp, t_0))$ depend on time through the difference $T \equiv t-t_0$.

We rewrite the Majorana density operator of Eq.(\ref{Eq:LeptonNumberDensity}) using the expansion with massless spinors in Eq.(\ref{eq:initialc});
\begin{multline}
  l^M_{\alpha}(t,\bx) = \int'\frac{d^3 \bk}{(2\pi)^3 2|\bk|}\int'\frac{d^3 \bp}{(2\pi)^3 2|\bp|}
\\
  \times\left[a^\dagger_\alpha(\bk,t)a_\alpha(\bp,t)\overline{u_L}(\bk)\gamma^0u_L(\bp)e^{-i(\bk-\bp)\cdot\bx}
  +b_\alpha(\bk,t)a_\alpha(\bp,t)\overline{v_L}(\bk)\gamma^0u_L(\bp)e^{i(\bk+\bp)\cdot\bx} \right.
\\
  +\left.a^\dagger_\alpha(\bk,t)b^\dagger_\alpha(\bp,t)\overline{u_L}(\bk)\gamma^0v_L(\bp)e^{-i(\bk+\bp)\cdot\bx}
  -b^\dagger_\alpha(\bp,t)b_\alpha(\bk,t)\overline{v_L}(\bk)\gamma^0v_L(\bp)e^{i(\bk-\bp)\cdot\bx}\right],
  \label{Eq:MajoranaDensity}
\end{multline}
where the integration region $\int'$ does not include the zero momentum mode.
Next, we can write the time evolution the Majorana density, solely in terms of the operators $a^{}_\alpha(\bp, t_0)$ and $b^{}_\alpha(\bp, t_0)$ by substituting the relations of Eq.(\ref{eq:timeDependOpa}) and Eq.(\ref{eq:timeDependOpb}).
The result for the time evolution of Eq.(\ref{Eq:LeptonNumber}) was first presented in Ref.\cite{Adam:2021qiq}.

To study the spacetime evolution, we take the expectation value of the Majorana density operator $\langle \psi_\sigma(q^0;\sigma_q), t_0 |l^M_{\alpha}(t,\bx)|\psi_\sigma(q^0;\sigma_q), t_0 \rangle$.
We assume the initial momentum state of $\lvert \psi_\sigma(q^0;\sigma_q), t_0 \rangle$ is sharply peaked at zero in the first and third components, but has a Gaussian distribution in the second, i.e., $\bq=(0,q,0)$.
This leads to a shape similar to a 1-D Gaussian wave packet,
\begin{equation}
  \lvert \psi_\sigma(q^0;\sigma_q), t_0 \rangle = \frac{1}{\sqrt{\sigma_q}(2\pi)^{3/4}}
  \int'\frac{dq}{\sqrt{A} \sqrt{2|q|}}e^{-\frac{(q-q^0)^2}{4\sigma_q^2}}
  a_\sigma^\dagger(\bq, t_0) \lvert 0(t_0) \rangle ; 
  \label{Eq:Wavepacket},
\end{equation}
where $A=(2\pi)^2\delta^2(0)$ denotes the area for the two-dimensional space $(x_1,x_3)$ perpendicular to the direction of the momentum.
This is a normalization factor that leads to  $\langle\psi_\sigma(q^0;\sigma_q), t_0 \rvert L_\sigma(t=t_0) \lvert \psi_\sigma(q^0;\sigma_q), t_0 \rangle=1$.
The vacuum  $\lvert 0(t_0)\rangle $ satisfies,
\begin{equation}
  a(\bp, t_0) \lvert 0(t_0) \rangle = b(\bp, t_0) \lvert 0(t_0)  \rangle=0 ,
\end{equation}
for all non-zero $\bp$.
In Eq.(\ref{Eq:Wavepacket}), $\sigma_q$ is the width of the Gaussian distribution in the second component of the momentum.
The second component of the mean momentum $q^0$ is positive for the Gaussian distribution. 
Sandwiching the Majorana density operator of Eq.(\ref{Eq:MajoranaDensity}) with Eq.(\ref{Eq:Wavepacket}) and taking the integration over $\bk$ and $\bp$ results in,
\begin{multline}
    \langle \psi_\sigma(q^0;\sigma_q), t_0 |l^M_{\alpha}(t,\bx)|\psi_\sigma(q^0;\sigma_q), t_0 \rangle = 
      \frac{1}{\sigma_q(2\pi)^{3/2}} \iint' \frac{dq'dq}{A}
      e^{-\frac{(q'-q^0)^2+(q-q^0)^2}{4\sigma_q^2}-i(q'-q)\mathbf{e}_2\cdot\bx} \\
    \times\left[\sum_{i,j} V^{\ast}_{\alpha i}V^{}_{\sigma i}V^{}_{\alpha j}V^{\ast}_{\sigma j}
     \left(\cos{E_i(q')T}+i\frac{|q'|}{E_i(q')}\sin{E_i(q')T}\right)
     \left(\cos{E_j(q)T}-i\frac{|q|}{E_j(q)}\sin{E_j(q)T}\right) \right. \\
    \left. -\sum_{i,j} V^{\ast}_{\alpha i}V^{\ast}_{\sigma i}V^{}_{\alpha j}V^{}_{\sigma j}
    \frac{m_j}{E_j(q')}\sin{E_j(q')T}\frac{m_i}{E_i(q)}\sin{E_i(q)T}\right];
    \label{Eq:kpExpectationValue}
\end{multline}
where $T=t-t_0$ is the time difference and $\mathbf{e}^T_2=(0,1,0)$ is the unit vector.
The integration over $x_1$ and $x_3$, results in a linear density, which we denote as
\begin{equation}
  \lambda^M_{\sigma\rightarrow\alpha}(T=t-t_0, x_2) \equiv 
  \iint dx_1 dx_3 \langle \psi_\sigma(q^0;\sigma_q), t_0|l^M_{\alpha}(t,\bx)|\psi_\sigma(q^0;\sigma_q), t_0\rangle.
\end{equation}
The result of the integration is,
\begin{multline}
  \lambda^M_{\sigma\rightarrow\alpha}(T=t-t_0, x_2)
  = \frac{1}{\sigma_q(2\pi)^{3/2}} \iint' {dq'dq} e^{-\frac{(q'-q^0)^2+(q-q^0)^2}{4\sigma_q^2}-i(q'-q)x_2} 
\\
  \times\left[\sum_{i,j} V^{\ast}_{\alpha i}V^{}_{\sigma i}V^{}_{\alpha j}V^{\ast}_{\sigma j}
    \left(\cos{E_i(q')T}+i\frac{|q'|}{E_i(q')}\sin{E_i(q')T}\right)
    \left(\cos{E_j(q)T}-i\frac{|q|}{E_j(q)}\sin{E_j(q)T}\right) \right. 
\\
  \left. -\sum_{i,j} V^{\ast}_{\alpha i}V^{\ast}_{\sigma i}V^{}_{\alpha j}V^{}_{\sigma j}
  \frac{m_j}{E_j(q')}\sin{E_j(q')T}\frac{m_i}{E_i(q)}\sin{E_i(q)T}\right];
  \label{Eq:LinearDensity}
\end{multline}
where $x_2=\mathbf{e}_2\cdot\bx$. We note that a factor of $1/A$ is absent in the linear density because $\iint dx_1 dx_3 (1/A)=1$.

To perform the integration over $q'$ and $q$ in Eq.(\ref{Eq:LinearDensity}) we must assume two properties about the Gaussian distributions.
First, the distributions are sharply peaked around the mean momentum value $q^0$, i.e., $\sigma_q \ll q^0$.
Second, the width (variance) of the distribution $\sigma_q$ does not change in spacetime.
Those two assumptions allow us to approximate the $q'$ and $q$ integration as Gaussian; because,
\begin{equation}
    E_{i,j}(q^{(\prime)}) \simeq E_{i,j}(q^0)+\frac{q^0}{E_{i,j}(q^0)}(q^{(\prime)}-q^0).
    \label{Eq:IntegralApprox}
\end{equation}
The Gaussian integration of Eq.(\ref{Eq:LinearDensity}) leads to,
\begin{multline}
  \lambda^M_{\sigma\rightarrow\alpha}(T=t-t_0,x_2) \simeq \frac{\sigma_q}{(2\pi)^{1/2}}\sum_{i,j} V^{\ast}_{\alpha i}V^{}_{\sigma i}V^{}_{\alpha j}V^{\ast}_{\sigma j}
\\
  \times\frac{1}{2}\left[(v_{i0}+v_{j0}+1+v_{i0}v_{j0})e^{i(E_i(q^0)-E_j(q^0))T }e^{-\sigma^2_q[(x_2-v_{i0}T)^2+(x_2-v_{j0}T)^2]}\right.
\\
  \quad\quad-(v_{i0}+v_{j0}-1-v_{i0}v_{j0})e^{-i(E_i(q^0)-E_j(q^0))T}e^{-\sigma^2_q[(x_2+v_{i0}T)^2+(x_2+v_{j0}T)^2]}
\\
  \quad\quad+(v_{i0}-v_{j0}+1-v_{i0}v_{j0})e^{i(E_i(q^0)+E_j(q^0))T}e^{-\sigma^2_q[(x_2-v_{i0}T)^2+(x_2+v_{j0}T)^2]}
\\
  \left.\quad\quad-(v_{i0}-v_{j0}-1+v_{i0}v_{j0})e^{-i(E_i(q^0)+E_j(q^0))T}e^{-\sigma^2_q[(x_2+v_{i0}T)^2+(x_2-v_{j0}T)^2]} \right] 
\\
  -\frac{\sigma_q}{(2\pi)^{1/2}}\sum_{i,j} V^{\ast}_{\alpha i}V^{\ast}_{\sigma i}V^{}_{\alpha j}V^{}_{\sigma j} \sqrt{1-v^2_{i0}}\sqrt{1-v^2_{j0}}
\\
  \times\frac{1}{2}\left[e^{i(E_i(q^0)-E_j(q^0))T}e^{-\sigma^2_q[(x_2-v_{i0}T)^2+(x_2-v_{j0}T)^2]} \right.
\\
  \quad+e^{-i(E_i(q^0)-E_j(q^0))T}e^{-\sigma^2_q[(x_2+v_{i0}T)^2+(x_2+v_{j0}T)^2]}
\\
  \quad-e^{i(E_i(q^0)+E_j(q^0))T}e^{-\sigma^2_q[(x_2-v_{i0}T)^2+(x_2+v_{j0}T)^2]}
\\
  \left.-e^{-i(E_i(q^0)+E_j(q^0))T}e^{-\sigma^2_q[(x_2+v_{i0}T)^2+(x_2-v_{j0}T)^2]} \right],
\label{Eq:ExpectationValue}
\end{multline}
where $v_{i,j0}=q^0/E_{i,j}(q^0)$ are the group velocities of the distributions. 
The terms with the PMNS matrix combination $V^{\ast}_{\alpha i}V^{\ast}_{\sigma i}V^{}_{\alpha j}V^{}_{\sigma j}$ are dependent on the Majorana phases and are suppressed by the small masses of neutrinos $\sqrt{1-v^2_{i,j0}}=m_{i,j}/E_{i,j}(q^0)$.
Lastly, to compare with our previous work's result that considered only the time evolution of plane waves, we integrate the linear density of Eq.(\ref{Eq:ExpectationValue}) over all space,
\begin{multline}
  \int \lambda^M_{\sigma\rightarrow\alpha}(T=t-t_0,x_2) dx_2 \simeq \frac{1}{4}\sum_{i,j}V^{\ast}_{\alpha i}V^{}_{\sigma i}V^{}_{\alpha j}V^{\ast}_{\sigma j}
\\
  \times\left[(v_{i0}+v_{j0}+1+v_{i0}v_{j0})e^{i(E_i(q^0)-E_j(q^0))T }e^{-\sigma^2_q\frac{(v_{i0}-v_{j0})^2T^2}{2}}\right.
\\
  \quad\quad-(v_{i0}+v_{j0}-1-v_{i0}v_{j0})e^{-i(E_i(q^0)-E_j(q^0))T}e^{-\sigma^2_q\frac{(v_{i0}-v_{j0})^2 T^2}{2}}
\\
  \quad\quad+(v_{i0}-v_{j0}+1-v_{i0}v_{j0})e^{i(E_i(q^0)+E_j(q^0))T}e^{-\sigma^2_q\frac{(v_{i0}+v_{j0})^2 T^2}{2}}
\\
  \left.\quad\quad-(v_{i0}-v_{j0}-1+v_{i0}v_{j0})e^{-i(E_i(q^0)+E_j(q^0))T}e^{-\sigma^2_q\frac{(v_{i0}+v_{j0})^2 T^2}{2}} \right]
\\
  -\frac{1}{4}\sum_{i,j}V^{\ast}_{\alpha i}V^{\ast}_{\sigma i}V^{}_{\alpha j}V^{}_{\sigma j}\sqrt{1-v^2_{i0}}\sqrt{1-v^2_{j0}}
\\
  \times\left[ (e^{i(E_i(q^0)-E_j(q^0))T}+e^{-i(E_i(q^0)-E_j(q^0))T}) e^{-\sigma^2_q\frac{(v_{i0}-v_{j0})^2 T^2}{2}}\right. 
\\
  \left.-(e^{i(E_i(q^0)+E_j(q^0))T}+e^{-i(E_i(q^0)+E_j(q^0))T})e^{-\sigma^2_q\frac{(v_{i0}+v_{j0})^2T^2}{2}} \right] .
\label{Eq:MajoranaSpaceTimeInt}
\end{multline}
The result of our previous work \cite{Adam:2021qiq} is recovered in the plane wave limit, for which we take the width $\sigma_q$ of the density in momentum space to zero.

\subsection{Dirac neutrino formulation}
For a Dirac neutrino, we start from the following Lagrangian,
\begin{equation}
	\mathcal{L}^D= \overline{\nu_{L\alpha}} i \gamma^\mu \partial_\mu \nu_{L\alpha}
	  +\overline{\nu_{R\alpha}} i \gamma^\mu \partial_\mu \nu_{R\alpha}
	  -\left(\overline{\nu_{R\alpha}}m_{\alpha\beta}\nu_{L\beta} 
    + \overline{\nu_{L \alpha}}(m^\dagger)_{\alpha\beta}\nu_{R\beta} \right).
	\label{eq:Lagmulti1}
\end{equation}
The first and second terms are kinetic ones and third term is Dirac mass one.
The equation of motion is derived as,
\begin{gather}
  i \gamma^\mu \partial_\mu  \nu_{L \alpha} =(m^\dagger)_{\alpha \beta} \nu_{R \beta}
  \label{eq:equmotfa}
\\
  i \gamma^\mu \partial_\mu  \nu_{R \alpha} = m_{\alpha \beta} \nu_{L \beta}.
  \label{eq:equmotfb}
\end{gather}
We expand left and right chiral fields by massless spinors
\begin{gather}
	\nu_{L\alpha}(t,\bx) = \int^\prime \frac{d^3\bp}{(2\pi)^3 2|\bp|} \left( u_L(\bp)a_{L\alpha}(\bp, t)e^{i \bp\cdot\bx} + v_L(\bp) b_{L\alpha}^\dagger(\bp, t)e^{-i\bp\cdot\bx} \right), \label{eq:leftmasslessexpand}
\\
    \nu_{R\alpha}(t,\bx) = \int^\prime \frac{d^3\bp}{(2\pi)^3 2|\bp|} \left( u_R(\bp)
a_{R\alpha}(\bp, t)e^{i\bp\cdot\bx} + v_R(\bp) b_{R\alpha}^\dagger(\bp, t)e^{-i\bp\cdot\bx} \right).
\label{eq:rightmasslessexpand}
\end{gather}
We have introduced a creation and annihilation operators for the fields for each chirality. 
The massless spinors for left chirality field $\nu_{L\alpha}$ are the same as Eq.(\ref{eq:spinormathn}), while the following massless spinors are used for right chirality field, 
\begin{equation}
  u_R(\bp)=-v_R(\bp)=\sqrt{\abs{2\bp}}\begin{pmatrix} \phi_+(\bn) \\ 0 \end{pmatrix}
  \label{eq:rightspinormathn}.
\end{equation}

In contrast to the Majorana case, the Dirac mass matrix is diagonalized by two mixing matrices,
\begin{gather}
    \nu_{L\beta}=V_{\beta j}\nu_{Lj},
  \\
    \nu_{R\alpha}=U_{\alpha i}\nu_{Ri},
  \\
    (U^\dagger)_{i\alpha}m_{\alpha\beta}V_{\beta j}=m_i\delta_{ij} .
\end{gather}
Then the equations of motion, Eq.(\ref{eq:equmotfa}) and Eq.(\ref{eq:equmotfb}), for the mass basis turn into the following form,
\begin{gather}
  i \gamma^\mu \partial_\mu  \nu_{L i} =m_{i} \nu_{R i} ,
\\
  i \gamma^\mu \partial_\mu  \nu_{R i} =m_{i} \nu_{L i} .
\end{gather}
The Dirac field with the definite mass $m_i$ is constructed as 
\begin{align}
  \psi_{D i}({\bx},t)&= \nu_{L i}({\bf x}, t)+ \nu_{R i}({\bf x}, t) 
  \label{eq:Dirac1}
\\
  &= V_{\alpha i}^\ast \nu_{L \alpha } ({\bf x}, t)+ U_{\alpha i}^\ast\nu_{R \alpha } ({\bf x}, t).
  \label{eq:Dirac2}
\end{align}
We expand the Dirac field in Eq.(\ref{eq:Dirac1}) using massive spinors,
\begin{equation}
  \psi_{D i}({\bx},t)=\int^\prime \frac{d^3\bp}{(2\pi)^3 2E_i(\bp)} \sum_{\lambda} \left(u_i(\bp,\lambda)a_i(\bp,\lambda)e^{-i (E_i t - \bp\cdot\bx) } + v_i(\bp,\lambda)b_i^\dagger(\bp,\lambda)e^{i (E_i t - \bp\cdot\bx)}\right),
  \label{eq:onshellDirac}
\end{equation}
where $\lambda$ is the helicity.
The mass operators $a_i(\bp,\lambda)$ and $b_i(\bp,\lambda)$ obey the usual anti-commutation relations,
\begin{align}
	\{a_i(\bp,\lambda), a_j^\dagger(\bq,\lambda)\} & = (2\pi)^3 2E_i(\bp) \delta_{ij}\delta^{(3)}(\bp-\bq), \label{eq:massOperatorsa}
\\
	\{b_i(\bp,\lambda), b_j^\dagger(\bq,\lambda)\} & = (2\pi)^3 2E_i(\bp) \delta_{ij}\delta^{(3)}(\bp-\bq). \label{eq:massOperatorsb}
\end{align}
All other anti-commutation relations are zero.
Using Eq.(\ref{eq:leftmasslessexpand}), Eq.(\ref{eq:rightmasslessexpand}), and Eqs.(\ref{eq:Dirac1}-\ref{eq:onshellDirac}), the operators for the flavor basis can be related to the mass basis.
\begin{align}
	\frac{1}{\sqrt{2|\bp|}}a_{L\alpha}(\pm \bp, t) & = \sum_{i}^3
      V_{\alpha i}\frac{\sqrt{E_i(\bp)+|\bp|}}{2E_i(\bp)}
      \left(a_i(\pm \bp,-)  e^{-i E_i(\bp) t} \pm i\frac{m_i}{E_i(\bp)+|\bp|}b_i^\dagger(\mp \bp,-)  e^{i E_i(\bp) t}\right), \label{eq:operatorarelations}
\\
	\frac{1}{\sqrt{2|\bp|}}b_{L\alpha}^\dagger(\pm \bp,t) & = \sum_{i}^3
      V_{\alpha i}\frac{\sqrt{E_i(\bp)+|\bp|}}{2E_i(\bp)}
      \left(b_i^\dagger(\pm \bp,+) e^{i E_i(\bp) t} \mp i\frac{m_i}{E_i(\bp)+|\bp|} 
a_i(\mp \bp,+)  e^{-i E_i(\bp) t}\right). \label{eq:operatorbrelations}
\end{align}
Where the relations for the right-handed operators $a_{R\alpha},\,b_{R\alpha}$ are found by replacing the PMNS matrices $V\rightarrow U$ and flipping the operator helicity $a_i(\pm \bp,\pm)\rightarrow a_i(\pm \bp,\mp)$, $b_i(\pm \bp,\pm)\rightarrow b_i(\pm \bp,\mp)$\footnote{The Dirac case operator relations first appeared in the thesis of \cite{Benoit:2022dsc}.}.
Again, the operators obey the usual anti-commutation relations of,
\begin{align}
	\{a_{L\alpha}(\pm \bp, t), a_{L\beta}^\dagger(\pm \bq, t)\} &= 2|\bp|(2\pi)^3\delta^{(3)}(\bp-\bq)\delta_{\alpha\beta}, \label{eq:flavorOperatorsa}
\\
	\{b_{L\alpha}(\pm \bp,t), b_{L\beta}^\dagger(\pm \bq,t)\} &= 2|\bp|(2\pi)^3\delta^{(3)}(\bp-\bq)\delta_{\alpha\beta}, \label{eq:flavorOperatorsb}
\end{align}
with all others being zero. Similar anti-commutation relations hold for the
right-handed operators.

The time evolution of Eq.(\ref{eq:operatorarelations}) and Eq.(\ref{eq:operatorbrelations}) starting from operators defined at 
$t=t_0$, can be calculated using a similar method to the Majorana case, Eq.(\ref{eq:timeDependOpa}) and Eq.(\ref{eq:timeDependOpb}).
We first write the massive operators $a^{}_{i}(\bp,\lambda)$ and $b^\dagger_{i}(\bp,\lambda)$ in terms of the operators $a_{L \alpha}(\bp, t_0)$, $b_{L \alpha}(\bp, t_0)$, $a_{R \alpha}(\bp, t_0)$, and $b_{R \alpha}(\bp, t_0)$ using the inverted relations of Eq.(\ref{eq:operatorarelations}), Eq.(\ref{eq:operatorbrelations}), and their right-handed counterparts.
Then we substitute those inverted relations at $t=t_0$ into the massive operators of Eq.(\ref{eq:operatorarelations}) and Eq.(\ref{eq:operatorbrelations}) to find the time evolution to be,
\begin{gather} 
  \begin{split}
    a_{L\alpha}(\pm\bp,t) = \sum_{i=1}^3\sum_{\beta=e}^\tau
     & \left[V^{}_{\alpha i}V^\ast_{\beta i}\left(\cos E_i(\bp)T - i\frac{|\bp|}{E_i(\bp)}\sin E_i(\bp)T \right)a_{L\beta}(\pm\bp,t_0) \right. \\
     & \left. \quad \mp V^{}_{\alpha i}U^\ast_{\beta i}\frac{m_i}{E_i(\bp)}\sin E_i(\bp)T \, b^\dagger_{R\beta}(\mp\bp,t_0) \right],
  \end{split} \\
  \begin{split}
    a^{\dagger}_{L\alpha}(\pm\bp,t) = \sum_{i=1}^3\sum_{\gamma=e}^\tau
     & \left[V^\ast_{\alpha i}V^{}_{\gamma i} \left(\cos E_i(\bp)T + i\frac{|\bp|}{E_i(\bp)}\sin E_i(\bp)T \right)a^\dagger_{L\gamma}(\pm\bp,t_0) \right. \\
     & \left. \quad\mp V^\ast_{\alpha i}U^{}_{\gamma i}\frac{m_i}{E_i(\bp)}\sin E_i(\bp)T \, b_{R\gamma}(\mp\bp,t_0) \right],
  \end{split}\\
  \begin{split}
    b_{L\alpha}(\pm\bp,t) = \sum_{i=1}^3\sum_{\beta=e}^\tau
     & \left[V^\ast_{\alpha i}V^{}_{\beta i} \left(\cos E_i(\bp)T - i\frac{|\bp|}{E_i(\bp)}\sin E_i(\bp)T \right)b_{L\beta}(\pm\bp,t_0) \right. \\
     & \left. \quad\mp V^\ast_{\alpha i}U^{}_{\beta i}\frac{m_i}{E_i(\bp)}\sin E_i(\bp)T \, a^\dagger_{R\beta}(\mp\bp,t_0) \right],
  \end{split}\\
  \begin{split}
    b^{\dagger}_{L\alpha}(\pm\bp,t) = \sum_{i=1}^3\sum_{\gamma=e}^\tau
     & \left[V^{}_{\alpha i}V^\ast_{\gamma i} \left(\cos E_i(\bp)T + i\frac{|\bp|}{E_i(\bp)}\sin E_i(\bp)T \right)b^\dagger_{L\gamma}(\pm\bp,t_0) \right. \\
     & \left. \quad\mp V^{}_{\alpha i}U^\ast_{\gamma i}\frac{m_i}{E_i(\bp)}\sin E_i(\bp)T \, a_{R\gamma}(\mp\bp,t_0) \right],
  \end{split}
\end{gather}
where the right-handed operators are just replacements of the PMNS matrices $U\rightarrow V$, $V\rightarrow U$ and the handedness $a_{(L,R)} \rightarrow a_{(R,L)}$, $b_{(L,R)} \rightarrow b_{(R,L)}$.
We define the lepton family number density, for both the left- and right-handed Dirac neutrinos,
\begin{gather}
	l^L_\alpha (t,\bx) = \,:\overline{\nu_{L\alpha}}(t,\bx)\gamma^0 \nu_{L\alpha}(t,\bx):\,,\label{Eq:left-hand_Lep}\\
	l^R_\alpha (t,\bx) = \,:\overline{\nu_{R\alpha}}(t,\bx)\gamma^0 \nu_{R\alpha}(t,\bx): \,. \label{Eq:right-hand_Lep}
\end{gather}
Then, the evolution of the lepton family number density for the left-handed case Eq.(\ref{Eq:left-hand_Lep}) is obtained by substituting the time dependent form of Eq.(\ref{eq:leftmasslessexpand}),
\begin{multline}
    l^L_{\alpha}(t,\bx) = \int'\frac{d^3 \bk}{(2\pi)^3 2|\bk|}\int'\frac{d^3 \bp}{(2\pi)^3 2|\bp|} \\
    \times\left[a^\dagger_{L\alpha}(\bk,t)a_{L\alpha}(\bp,t)\overline{u_L}(\bk)\gamma^0u_L(\bp)e^{-i(\bk-\bp)\cdot\bx}
    +b_{L\alpha}(\bk,t)a_{L\alpha}(\bp,t)\overline{v_L}(\bk)\gamma^0u_L(\bp)e^{i(\bk+\bp)\cdot\bx} \right.\\
    +\left.a^\dagger_{L\alpha}(\bk,t)b^\dagger_{L\alpha}(\bp,t)\overline{u_L}(\bk)\gamma^0v_L(\bp)e^{-i(\bk+\bp)\cdot\bx}
    -b^\dagger_{L\alpha}(\bp,t)b_{L\alpha}(\bk,t)\overline{v_L}(\bk)\gamma^0v_L(\bp)e^{i(\bk-\bp)\cdot\bx}\right],
    \label{Eq:DiracDensity}
\end{multline}
whereas the right-handed form is a replacement of the spinors and operators from $L$ to $R$.
Next, we write the expectation value using a similar state $|\psi_\sigma(q^0;\sigma_q), t_0 \rangle$ as Eq.(\ref{Eq:Wavepacket}), but with the left-handed operator replaced as $a^\dagger_{\sigma}(\bq, t_0)\rightarrow a^\dagger_{L\sigma}(\bq, t_0)$.
We sandwich the left-handed density for the lepton family number of Eq.(\ref{Eq:DiracDensity}) with the left-handed state $\vert\psi^L_\sigma(q^0;\sigma_q), t_0 \rangle$ and integrate over $\bk,\bp$ to result in,
\begin{multline}
    \langle \psi^L_\sigma(q^0;\sigma_q) ,t_0 \lvert l^L_{\alpha}(t,\bx)\rvert \psi^L_\sigma(q^0;\sigma_q), t_0 \rangle = 
      \frac{1}{\sigma_q(2\pi)^{3/2}} \iint' \frac{dq'dq}{A}
      e^{-\frac{(q'-q^0)^2+(q-q^0)^2}{4\sigma_q^2}-i(q'-q)\mathbf{e}_2\cdot\bx} \\
    \times\left[\sum_{i,j} V^{\ast}_{\alpha i}V^{}_{\sigma i}V^{}_{\alpha j}V^{\ast}_{\sigma j}
     \left(\cos{E_i(q')T}+i\frac{|q'|}{E_i(q')}\sin{E_i(q')T}\right)
     \left(\cos{E_j(q)T}-i\frac{|q|}{E_j(q)}\sin{E_j(q)T}\right)\right] .
    \label{Eq:DiracExpectationValue}
\end{multline}
An initial right-handed state of Eq.(\ref{Eq:DiracExpectationValue}), $\langle \psi^R_\sigma(q^0;\sigma_q), t_0 \lvert l^R_{\alpha}(t,\bx)\rvert\psi^R_\sigma(q^0;\sigma_q), t_0 \rangle$, is obtained by interchanging the matrices as $U\rightarrow V,\quad V\rightarrow U$.
Lastly, we perform the integration over $q$ and $q'$ using the approximation of Eq.(\ref{Eq:IntegralApprox}) to write the integrand in Gaussian form.
The result is the linear density expectation value of the left-handed lepton family number,
\begin{equation}
    \begin{split}
      \lambda^L_{\sigma\rightarrow\alpha}(T=t-t_0,x_2)\simeq & \frac{\sigma_q}{(2\pi)^{1/2}}\sum_{i,j} V^{\ast}_{\alpha i}V^{}_{\sigma i}V^{}_{\alpha j}V^{\ast}_{\sigma j} \\
      &\times\frac{1}{2}\left[ (v_{i0}+v_{j0}+1+v_{i0}v_{j0})e^{i(E_i(q^0)-E_j(q^0))T}e^{-\sigma^2_q[(x_2-v_{i0}T)^2+(x_2-v_{j0}T)^2]}\right. \\
      & \quad\quad -(v_{i0}+v_{j0}-1-v_{i0}v_{j0})e^{-i(E_i(q^0)-E_j(q^0))T}e^{-\sigma^2_q[(x_2+v_{i0}T)^2+(x_2+v_{j0}T)^2]} \\
      & \quad\quad +(v_{i0}-v_{j0}+1-v_{i0}v_{j0})e^{i(E_i(q^0)+E_j(q^0))T}e^{-\sigma^2_q[(x_2-v_{i0}T)^2+(x_2+v_{j0}T)^2]} \\
      & \left.\quad\quad -(v_{i0}-v_{j0}-1+v_{i0}v_{j0})e^{-i(E_i(q^0)+E_j(q^0))T}e^{-\sigma^2_q[(x_2+v_{i0}T)^2+(x_2-v_{j0}T)^2]} \right].
    \end{split}
     \label{Eq:LeftHandedExpectationValue}
\end{equation}
Where we have defined $\lambda^L_{\sigma\rightarrow\alpha}(T=t-t_0,x_2)=
\iint dx_1 dx_3 \langle \psi^L_\sigma(q^0;\sigma_q), t_0|l^L_{\alpha}(t,\bx)|\psi^L_\sigma(q^0;\sigma_q), t_0 \rangle$.
The process for solving for the linear density of the right-handed lepton family number of Eq.(\ref{Eq:right-hand_Lep}) is the same.
After sandwiching the right-handed density of Eq.(\ref{Eq:right-hand_Lep}) with the left-handed state $\vert\psi^L_\sigma(q^0;\sigma_q), t_0\rangle$ and integrating over $\bk,\bp$ our result is,
\begin{multline}
    \langle \psi^L_\sigma(q^0;\sigma_q),t_0|l^R_{\alpha}(t,\bx)|\psi^L_\sigma(q^0;\sigma_q),t_0\rangle = 
    \frac{1}{\sigma_q(2\pi)^{3/2}} \iint' \frac{dq'dq}{A}
    e^{-\frac{(q'-q^0)^2+(q-q^0)^2}{4\sigma_q^2}-i(q'-q)\mathbf{e}_2\cdot\bx} \\
  \times\left[\sum_{i,j} U^{}_{\alpha j}V^{\ast}_{\sigma j}U^{\ast}_{\alpha i}V^{}_{\sigma i}
  \frac{m_j}{E_j(q')}\sin{E_j(q')T}\frac{m_i}{E_i(q)}\sin{E_i(q)T}\right].
\end{multline}
Thus, we write the resulting linear density expectation value of the right-handed lepton family number,
\begin{multline}
    \lambda^R_{\sigma\rightarrow\alpha}(T=t-t_0,x_2) 
\simeq \frac{\sigma_q}{\sqrt{2\pi}}\sum_{i,j}\sqrt{1-v_{i0}^2} \sqrt{1-v_{j0}^2} e^{-\sigma_q^2\left((v_{i0}^2+v_{j0}^2)T^2+2 x_2^2\right)} \\
    \times \left\{\text{Re}\left(U^{\ast}_{\alpha i}V^{}_{\sigma i}U^{}_{\alpha j}V^{\ast}_{\sigma j} \right)
    \left[\cosh\left(2\sigma_q^2(v_{i0}+v_{j0}) T x_2\right)\cos\left((E_i(q^0)-E_j(q^0))T\right)\right.\right. \\
    \left.-\cosh\left(2\sigma_q^2(v_{i0}-v_{j0}\right)T x_2)\cos\left((E_i(q^0)+E_j(q^0))T\right) \right] \\
    -\text{Im}\left(U^{\ast}_{\alpha i}V^{}_{\sigma i}U^{}_{\alpha j}V^{\ast}_{\sigma j} \right)\left[\sinh\left(2\sigma_q^2(v_{i0}+v_{j0})Tx_2\right)\sin\left((E_i(q^0)-E_j(q^0))T\right)\right.\\
    \left.\left.-\sinh\left(2\sigma_q^2(v_{i0}-v_{j0})Tx_2\right)\sin\left((E_i(q^0)+E_j(q^0))T\right) \right] \right\},
     \label{Eq:RightHandedExpectationValue}
\end{multline}
where $\lambda^R_{\sigma\rightarrow\alpha}(T=t-t_0,x_2) = \iint dx_1 dx_3 \langle \psi^L_\sigma(q^0;\sigma_q),t_0|l^R_{\alpha}(t,\bx)|\psi^L_\sigma(q^0;\sigma_q), t_0\rangle$.

\subsection{Comparison to Wave packet formulations}\label{sec:Decoherence}
The results we present in Eqs.(\ref{Eq:ExpectationValue}, \ref{Eq:LeftHandedExpectationValue}, \ref{Eq:RightHandedExpectationValue}) are based on the evolution of a density operator.
We will now discuss an interpretation of those results based on wave packet formulations.
Specifically, wave packets have been used in connection with neutrino oscillation phenomenology (see Ref. \cite{Beuthe:2001rc} for a review) as several forms; the quantum mechanical (QM) formulation \cite{Giunti:1991ca}, and the external wave packet QFT formulation \cite{Giunti:1993se,Giunti:2002xg}.
We will focus on comparisons to interpretations in the quantum mechanical formulation, because it is the simplest in literature.

Often discussed in literature see \cite{Akhmedov:2019iyt,Kayser:1981ye,Akhmedov:2017xxm,Akhmedov:2009rb}, neutrino oscillations occur when three types of coherence are satisfied.
\begin{enumerate}
    \item The different massive neutrino components are coherently produced.
    \item Coherent propagation of massive neutrino components.
    \item Coherent detection of the different massive components.
\end{enumerate}
We do not consider any production or detection processes, so we can not discuss the possibility of coherent detection in our formulation.
The initial density state of Eq.(\ref{Eq:Wavepacket}) captures the properties of coherent propagation, similar to the QM wave packet.

A prediction of the QM wave packet formulation is the coherence length $L^{\text{coh}}_{i,j}$ for oscillations \cite{Nussinov:1976uw}.
The coherence length is a measure of the group velocities for the mass eigenstates that appears after integration over time.
It acts as a damping factor on the oscillations in space.
We find a similar damping in our density formulation.
The damping appears as a real exponential component that depends quadratically on spacetime,
\begin{gather}
    e^{-\sigma^2_q[(x_2\pm v_{i0}T)^2+(x_2\pm v_{j0}T)^2]} \label{Eq:DampingLinearSquared1} \\
    e^{-\sigma^2_q[(x_2\pm v_{i0}T)^2+(x_2\mp v_{j0}T)^2]}. \label{Eq:DampingLinearSquared2}
\end{gather}
These real components are a non-linear damping that is applied to the density oscillation.
An important distinction between our density formulation and the QM formulation are the types of damping.
The QM formulation of Eq.(3) in Ref. \cite{Giunti:1991ca} has a single type of damping to coincide with the single type of oscillation
\begin{equation}
    e^{-\sigma^2_q[(x_2-v_{i0}T)^2+(x_2-v_{j0}T)^2]},
\end{equation}
which we denote as $(-,-)$.
Whereas, our density formulation has four different types damping, $(-,-),(-,+),(+,-),(+,+)$.
Each damping is applied to a different oscillation in Eqs.(\ref{Eq:ExpectationValue}, \ref{Eq:LeftHandedExpectationValue}, \ref{Eq:RightHandedExpectationValue}).
The $(+,+)$ is the strongest damping factor and $(-,-)$ is the weakest damping, which we illustrate by minimizing the polynomials inside the exponentials assuming $x_2>0$;
\begin{align}
    P_1\left(x_2,T=\frac{2x_2}{v_{i0}+v_{j0}}\right) & = 2x_2^2\frac{(v_{i0}-v_{j0})^2}{(v_{i0}+v_{j0})^2} &&\text{from }(-,-), 
    \label{eq:p1min} \\
    P_2\left(x_2,T=\frac{2x_2}{v_{i0}+v_{j0}}\right) & = 2x_2^2\frac{3(v_{i0}+v_{j0})^2+2(v_{i0}^2+v_{j0}^2)}{(v_{i0}+v_{j0})^2} && \text{from }(+,+), \\
    P_3\left(x_2,T=\frac{2x_2}{v_{i0}+v_{j0}}\right) & = 2x_2^2+8x_2^2\frac{v_{j0}^2}{(v_{i0}+v_{j0})^2} && \text{from }(-,+), \\
    P_4\left(x_2,T=\frac{2x_2}{v_{i0}+v_{j0}}\right) & = 2x_2^2+8x_2^2\frac{v_{i0}^2}{(v_{i0}+v_{j0})^2} && \text{from }(+,-).
\end{align}
Notice the peak value of $(-,-)$ is damped at the peak by $P_1<2x_2^2$ causing it to be weak.
Whereas the peak values of $(+,+)$, $(-,+)$, and $(+,-)$ are damped with $P_{2,3,4}>2x_2^2$ strengthening their suppression of the oscillations.

If we perturb slightly away from the peak time, the strength of the damping factors depend directly on the velocities.   For example,
\begin{equation}
    P_1\left(x_2,T=\frac{2x_2}{v_{i0}+v_{j0}}+\Delta T \right)-P_1\left(x_2,T=\frac{2x_2}{v_{i0}+v_{j0}}\right) = (v_{i0}^2+v_{j0}^2)(\Delta T)^2+\frac{2x_2(v_{i0}-v_{j0})^2}{v_{i0}+v_{j0}}\Delta T .
  \label{eq:p1perturbation}
\end{equation}
So, a decrease in the velocities lead to longer damping times. 

After we integrate over all space to arrive at Eq.(\ref{Eq:MajoranaSpaceTimeInt}), the damping factors appear as two types $\pm$.
\begin{equation}
    T^{coh}_{i,j}\simeq \frac{\sqrt{2}}{\sigma_q(v_{i0}\pm v_{j0})},
\end{equation}
which we write as a coherence time $T_{i,j}^{coh}$.
In the usual QM case only a single damping factor appears for the oscillations with $(v_{i0} -v_{j0})$.
However, new oscillation terms appear in our formulation corresponding to $e^{\pm i[E_i(q^0)+E_j(q^0)]T}$, to which a new damping factor with $(v_{i0} + v_{j0})$ is applied.

\section{Comparison of the Dirac and Majorana formulations}\label{sec:DiracVsMajorana}
From the expectation value equations of Eq.(\ref{Eq:ExpectationValue}), Eq.(\ref{Eq:LeftHandedExpectationValue}), and Eq.(\ref{Eq:RightHandedExpectationValue}) we find signatures of the Dirac and Majorana formulations.
Firstly, the terms proportional to the PMNS matrix combination of $V_{\alpha i}^\ast V^{}_{\sigma i}V^{}_{\alpha j} V_{\sigma j}^\ast$ are common between the formulations.
Those terms are the first summation from Eq.(\ref{Eq:ExpectationValue}) of the Majorana formulation and the left-handed Dirac formulation Eq.(\ref{Eq:LeftHandedExpectationValue}).
The difference between the formulations is the secondary term of the Majorana formulation and the right-handed Dirac formulation Eq.(\ref{Eq:RightHandedExpectationValue}).
The secondary term of the Majorana formulation has the PMNS combination $V^{\ast}_{\alpha i}V^{\ast}_{\sigma i}V_{\alpha j}V_{\sigma j}$ that is related to the Majorana phases and is subtracted from the common oscillation terms.
Whereas, the second unitary matrix $U_{\alpha i}$ from the right-handed Dirac formulation is present in Eq.(\ref{Eq:RightHandedExpectationValue}) and is added to Eq.(\ref{Eq:LeftHandedExpectationValue}) as $\lambda^D_{\sigma\rightarrow\alpha}(T=t-t_0 ,x_2)=\iint dx_1 dx_3 \langle \psi^L_\sigma(q^0;\sigma_q),t_0|l^L_{\alpha}(t,\bx)+l^R_{\alpha}(t,\bx)|\psi^L_\sigma(q^0;\sigma_q),t_0 \rangle$.
The sign of the different terms, subtraction for Majorana and addition for Dirac, is responsible for total lepton number violation in the Majorana case.
To help emphasize the affect of the sign difference we write the expectation value of the total lepton number for the Dirac formulation case,
\begin{equation}
	\sum_{\alpha} \lambda^D_{\sigma\rightarrow\alpha}(T=t-t_0,x_2) \simeq 
      \frac{\sigma_q}{\sqrt{2\pi}}\sum_i |V_{\sigma i}|^2\left[(1+v_{i0})e^{-2\sigma_q^2(x_2-v_{i0}T)^2}+(1-v_{i0})e^{-2\sigma_q^2(x_2+v_{i0}T)^2}\right],
\label{eq:DiracTL}
\end{equation}
where integration over $x_2$ would result in total lepton number conservation for all times.
Next, from Eq.(\ref{Eq:ExpectationValue}), we also take the summation over the lepton family numbers $\alpha$ as follows,
\begin{equation}
  \begin{split}
    \sum_{\alpha}\lambda^M_{\sigma\rightarrow\alpha}(T=t-t_0,x_2) \simeq & \frac{\sigma_q}{\sqrt{2\pi}}\sum_i | V_{\sigma i}|^2\left[v_{i0}(1+ v_{i0})e^{- 2\sigma_q^2 (x_2-v_{i 0} T)^2}-v_{i0} (1-v_{i0})e^{- 2\sigma_q^2 (x_2+v_{i 0} T)^2} \right.\\
    & \left. \hphantom{XXXXXXXXXX} +2(1-v_{i0}^2) e^{- 2\sigma_q^2 (v_{i 0}^2 T^2 + x_2^2)}\cos 2E_i(q^0)T \right].
  \end{split} \label{eq:MajoranaTL}
\end{equation}
In Eq.(\ref{eq:DiracTL}) for the Dirac case, there is no oscillation term and the linear density of the total lepton number is always positive.
For a nonrelativistic Majorana neutrino ($v_{i0} \ll 1$) in Eq.(\ref{eq:MajoranaTL}), the oscillation term can dominate and the sign of the linear density can change with respect to time.
\section{Illustration of nonrelativistic differences}

%

\begin{figure}[htb]
    \includegraphics[width=\textwidth]{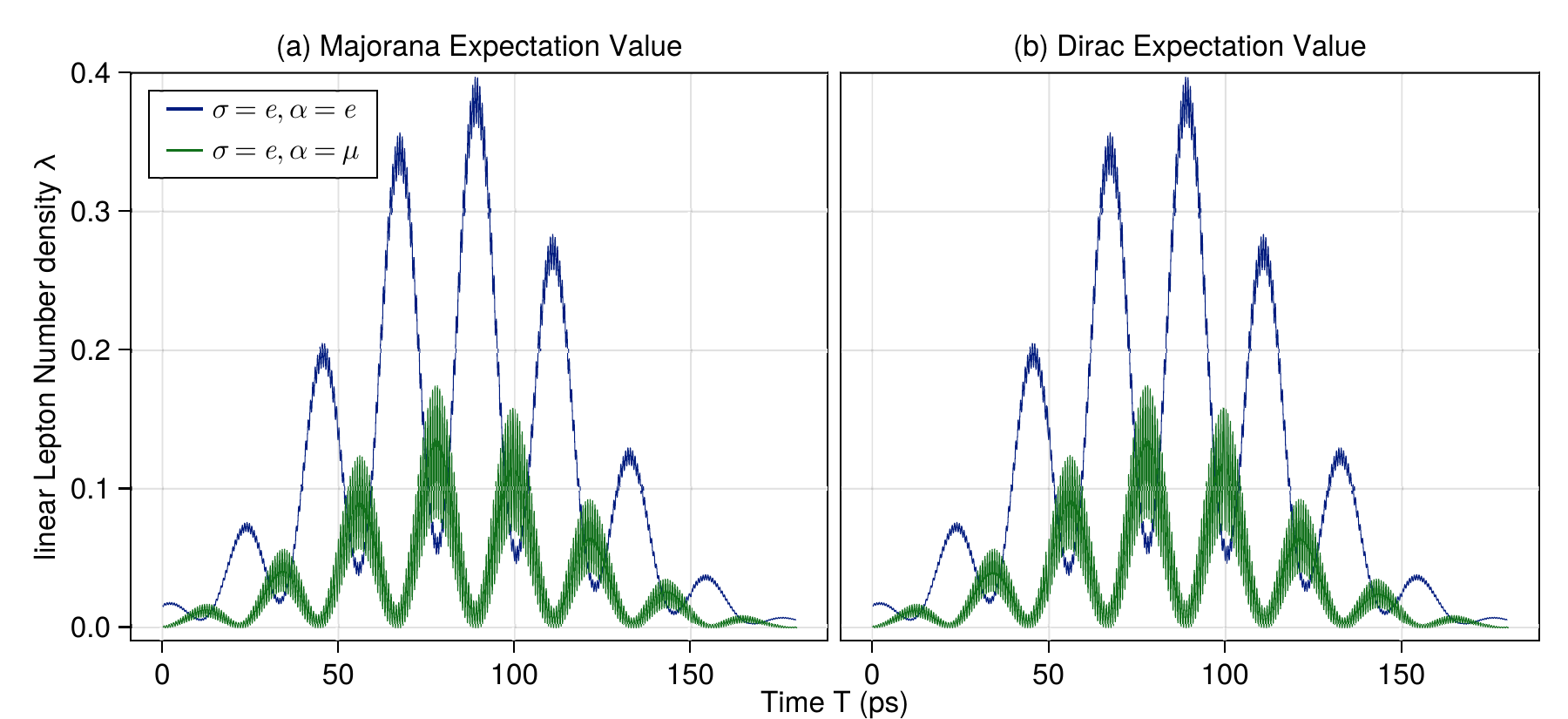}
    \caption{\label{Fig:3Flavor_Density_02e-mu} Time evolution of the linear density for the expectation value of a lepton family number with Majorana, panel (a), or Dirac, panel (b), mass.
    We take an arbitrary distance slice at $x_2=2.5$cm.
    The initial momentum density is a Gaussian distribution with a width of $\sigma_q = 0.00001$ and a mean momentum of $q^0=0.2$eV.
    Normal mass hierarchy is considered.
    Oscillation parameters are the best fit values from the NuFIT 5.0 (2020) collaboration \cite{Esteban:2020cvm}.}
\end{figure}

\begin{figure}[htb]
  \includegraphics[width=\textwidth]{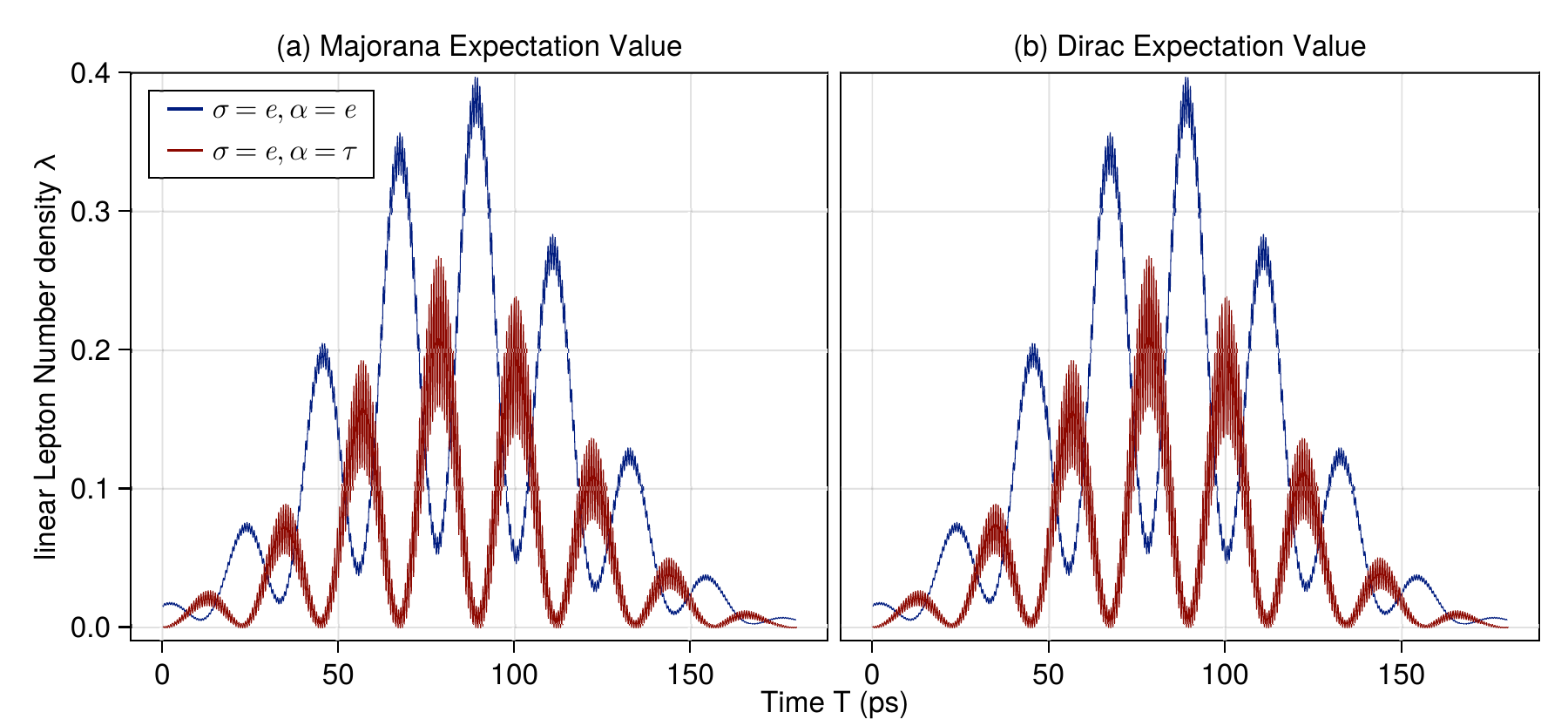}
  \caption{\label{Fig:3Flavor_Density_02e-tau} Similar to Fig.\ref{Fig:3Flavor_Density_02e-mu}, time evolution of the linear density for the expectation value of a lepton family number at $x_2 = 2.5$cm.
  However, we now consider the $\sigma=e,\,\alpha=\tau$ lepton family number values.
  The initial momentum density is a Gaussian distribution with a width of $\sigma_q = 0.00001$ and a mean momentum of $q^0=0.2$eV.
  Normal mass hierarchy is considered.
  Oscillation parameters are the best fit values from the NuFIT 5.0 (2020) collaboration \cite{Esteban:2020cvm}.}
\end{figure}

From inspection of the expectation values Eq.(\ref{Eq:ExpectationValue}), Eq.(\ref{Eq:LeftHandedExpectationValue}), and Eq.(\ref{Eq:RightHandedExpectationValue}); and the discussions in sections \ref{sec:Decoherence} and \ref{sec:DiracVsMajorana} we see two key results.
\begin{itemize}
    \item Differences between our formulation and the usual QM formulation are negligible under the ultra-relativistic assumption.
    \item The neutrino masses $\frac{m_i m_j}{E_i E_j}=\sqrt{1-v_i^2}\sqrt{1-v_j^2}$ suppress any modifications by the Dirac or Majorana mass formulations.
\end{itemize}
To illustrate those results we choose $m_{\text{lightest}}=0.01\text{ eV} < q^0 = 0.2\text{ eV}$ for the linear density expectations of Eqs.(\ref{Eq:ExpectationValue}) and (\ref{Eq:LeftHandedExpectationValue}).
Even for a momentum an order of magnitude greater than the mass, differences between the Majorana and Dirac cases are not visible (Fig. \ref{Fig:3Flavor_Density_02e-mu} and Fig. \ref{Fig:3Flavor_Density_02e-tau}).
We used the best fit values of the PMNS mixing angles $\theta_{12},\theta_{23}, \text{and }\theta_{13}$, the CP violating phase $\delta_{\text{CP}}$, and the mass squared differences $\Delta m_{21}^2 \text{ and } \Delta m_{31}^2$ from the work of the NuFIT 5.0 (2020) collaboration \cite{Esteban:2020cvm}.
Then, we are left to choose the absolute mass of the lightest neutrino $m_1 \text{ or } m_3$, the neutrino mass hierarchy,  normal $m_1<m_2\ll m_3$ or inverted $m_3\ll m_1<m_2$, the value of the mean momentum $q$, the width of the initial Gaussian density $\sigma_q$, the time $T=t-t_0$, the distance $x_2$, and the Majorana phases $\alpha_{21} \text{ and } \alpha_{31}$.
The definition of the Majorana phases is given as $\alpha_{21}=2\arg(V_{e2})$ and $\alpha_{31}=2\arg(V_{\mu 3})$.
Any changes caused by different Majorana phases are also hidden, because the differences between the Majorana and Dirac cases are not visible for $m_{\text{lightest}}=0.01\text{ eV} < q^0 = 0.2\text{ eV}$.

\begin{figure}[htb]
    \includegraphics[width=\textwidth]{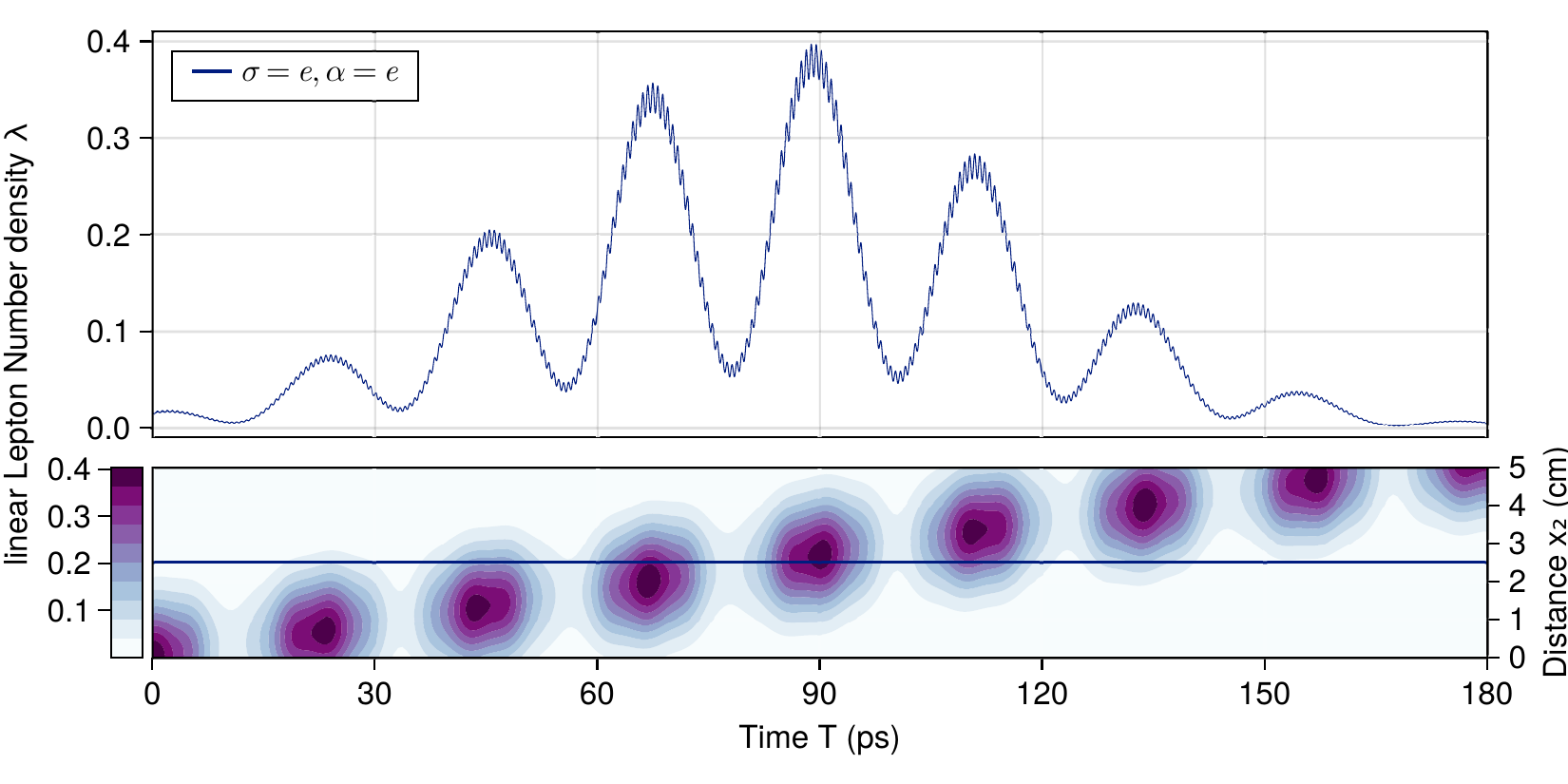}
    \caption{\label{Fig:DiracContour02} 2-D Spacetime contour of the linear density for the expectation value of the electron family number with a Dirac mass.
    The initial momentum density is a Gaussian distribution with a width of $\sigma_q = 0.00001$ and a mean momentum of $q^0=0.2$eV.
    Normal mass hierarchy for neutrinos is considered.
    Lepton mixing angles, the Dirac CP phase, and the mass squared differences are the reported best fit values from the work of the NuFIT 5.0 (2020) collaboration \cite{Esteban:2020cvm}.}
\end{figure}

The decoherence effects discussed in section \ref{sec:Decoherence} from the real exponentials of Eq.(\ref{Eq:ExpectationValue}), Eq.(\ref{Eq:LeftHandedExpectationValue}), and Eq.(\ref{Eq:RightHandedExpectationValue}) cause the oscillations to be localized to a spacetime region.
In Fig. \ref{Fig:DiracContour02}, we show the  2-D spacetime contour for an electron number linear density for the $\sigma = e \rightarrow \alpha = e$ case.  
The region where the electron number linear density has a maximum peak of $\sim 0.4$ propagates from the spacetime origin $(t,x_2)=(0,0)$ toward the upper right corner at a constant velocity.
The initial Gaussian shape is maintained during propagation because we assumed no wave packet spreading in the approximation of Eq.(\ref{Eq:IntegralApprox}).

\begin{figure}[htb]
   \includegraphics[width=\textwidth]{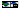}
    \caption{\label{Fig:3Flavor_Density_00002e-mu} Time evolution of the linear density for the expectation value of a lepton family number at $x_2 = 2.5$cm.
    We now consider the initial momentum density to have a mean momentum of $q^0=0.0002$eV for a Gaussian distribution with a width of $\sigma_q = 0.00001$.
    Compared to the preceding figures, differences between Majorana and Dirac are visible because $q^0 = 0.0002 < m_{lightest}=0.01$eV.
    Normal mass hierarchy is considered.
    Oscillation parameters are the reported best fit values from the work of the NuFIT 5.0 (2020) collaboration \cite{Esteban:2020cvm}.
    In addition, we choose the Majorana phases to be arbitrary values of $\alpha_{21}=\pi$ and $\alpha_{31}=0.5\pi$.}
\end{figure}

We discussed in section \ref{sec:DiracVsMajorana} how the total lepton number becomes violated in the Majorana case, and we illustrate that by decreasing the momentum to $ q^0 = 0.0002 < m_{lightest}=0.01$eV in Fig.\ref{Fig:3Flavor_Density_00002e-mu}.
This leads to the distinction between the Majorana and Dirac cases, in which the Majorana case has negative expectation values.
Both cases feature the small period oscillations of the inset graph because of lines 4 and 5 from Eq.(\ref{Eq:LeftHandedExpectationValue}).
Those lines are not present in the usual QM formulation, so these small period oscillations can distinguish our formulation.

By comparing Fig.\ref{Fig:3Flavor_Density_02e-mu} and Fig.\ref{Fig:3Flavor_Density_00002e-mu}, the oscillations occur for longer times when the average momentum is smaller.
This was discussed in section \ref{sec:Decoherence}, where Eq.(\ref{eq:p1perturbation}) was given as an example.
To recall, the strength of the damping factors depend directly on the velocities.  So, a decrease in the velocities lead to longer damping times.

\begin{figure}[htb]
    \includegraphics[width=\textwidth]{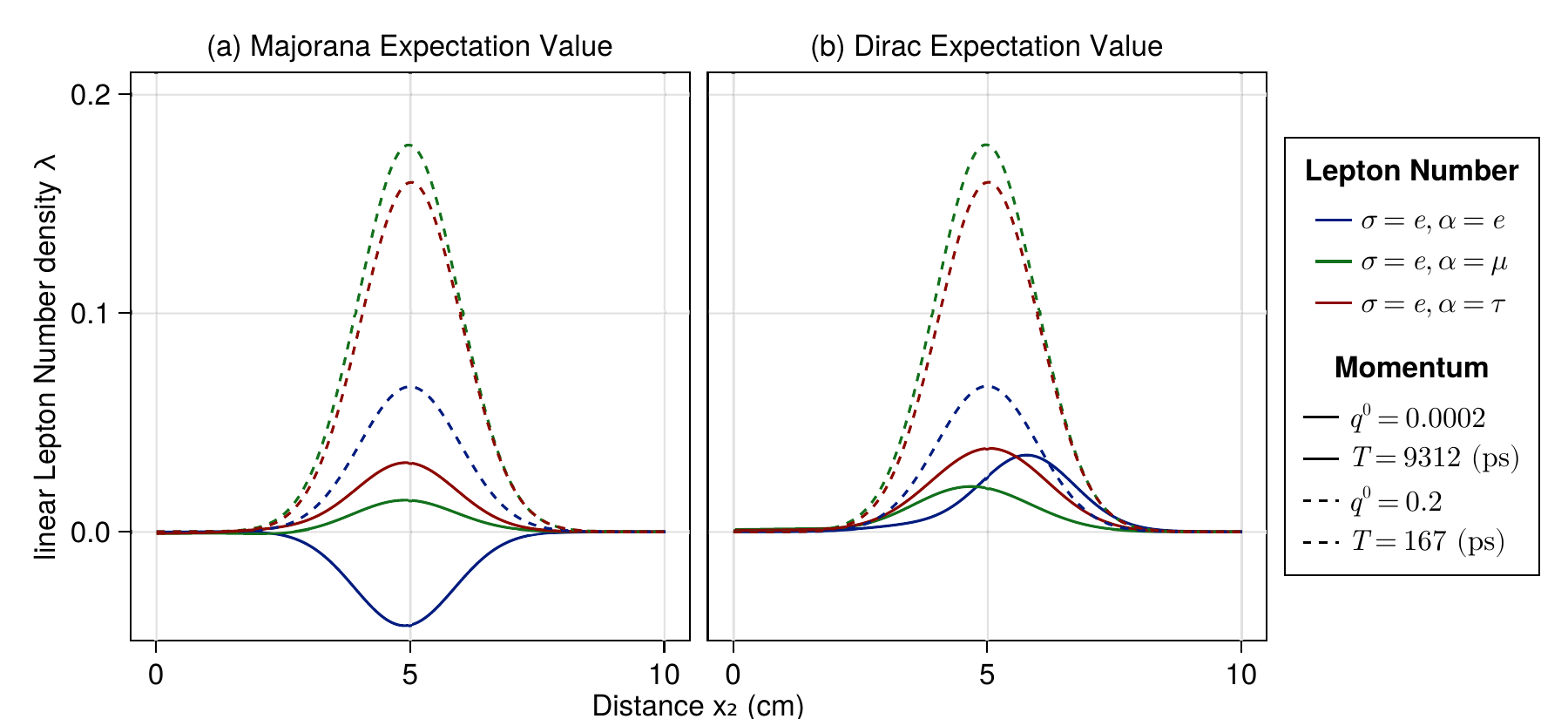}
    \caption{\label{Fig:3Flavor_Distance} Distance evolution of the linear density for the expectation value of a lepton family number.
    We have taken different time slices corresponding to when the peak of the linear densities are near $x_2=5.0$cm.
    Normal mass hierarchy is considered, and we choose the Majorana phases to be arbitrary values of $\alpha_{21}=\pi$ and $\alpha_{31}=0.5\pi$.
    Oscillation parameters are the reported best fit values from the work of the NuFIT 5.0 (2020) collaboration \cite{Esteban:2020cvm}.}
\end{figure}

Next, in the distance plane of Fig.\ref{Fig:3Flavor_Distance} the smaller momenta suppress the peak value for the linear density of the lepton family numbers.
This follows directly from Eq.(\ref{eq:p1min}), where for $q^0=0.2$ eV the polynomial is larger than for $q^0=0.0002$ eV.
Thus, the propagation coherence occurs at smaller distances, which is sometimes stated as a smaller coherence length.
Even with this in mind, at smaller momenta the Dirac and Majorana cases are distinguishable.  Similar to Fig.\ref{Fig:3Flavor_Density_00002e-mu}, the Majorana case has negative expectation values and the Dirac case is always positive.

\section{Conclusion}
We have proposed a novel formulation for neutrino oscillations from a QFT point of view.
This formulation can be applied to both relativistic and nonrelativistic energies for neutrinos.  
The previous formulation was limited to  Majorana neutrinos with a fixed momentum.
We have extended  the  previous formulation to the Dirac neutrino case and have applied to the case with a momentum distribution. 
To study the oscillation of neutrino with a momentum distribution, we have considered the linear lepton number density and the initial state with a Gaussian distribution along a one spacial direction. 
After taking the expectation value of the density with the state, we find a result similar to the quantum mechanical wave packet approach used to describe neutrino flavor oscillations.
Our result has additional terms that result in interesting behavior. 
In particular, we have studied type of decoherence ascribed to wave packet separation.
We have found  how the peak value of lepton number density is suppressed as it propagates.   
Even with that decoherence our formulation can distinguish between the neutrino mass type at nonrelativistic energies.
The mass types, the expectation value of the lepton number density for Majorana neutrinos can take both positive and negative value while for Dirac neutrinos, it does not change the sign.

\begin{acknowledgments}
The work of T.M. is supported by Japan Society for the Promotion of Science (JSPS) KAKENHI Grant Number JP17K05418.
We would like to thank Uma Sankar and Koichi Hamaguchi for useful comments and suggestions during the initial presentations of this work.
We would like to thank Tomohiro Inagaki, Tomoharu Orimi and Naoki Uemura for useful discussions.
Lastly, we thank the three anonymous referees for the detailed review and improvements.
\end{acknowledgments}

\appendix
\section{Quantization with the zero mode and the massless spinors}\label{sec:zeromodes}
In this appendix, we demonstrate the quantization with the massless spinor in Eq.(\ref{eq:initialc}) by adding the zero mode.
The zero mode corresponds a massive fermion in its rest frame. 
The quantization based on the Dirac bracket is carried out.
For simplicity, we consider only the single flavor version of the Lagrangian in Eq.(\ref{eq:LagMajoranamulti1}).
We write the Lagrangian  in terms of two component spinor $\eta$ including the zero mode as will be shown in Eq.(\ref{eq:cexpansion}),
\begin{equation}
  \nu_L(x,t)=\left(\begin{matrix} 0 \\ \eta(x,t) \end{matrix}\right).
  \label{eq:psiL}
\end{equation}
Then the Lagrangian is given as follows,
\begin{equation}
  \mathcal{L} = \eta^\dagger i \overline{\sigma}^\mu \partial_\mu \eta - \frac{m}{2}\left(-\eta^\dagger i\sigma_2 \eta^\ast + \eta^T i\sigma_2 \eta \right).
\end{equation}
By treating $\eta^\dagger$ as independent variables, 
\begin{align}
  \pi_\eta = \frac{\partial \mathcal{L}}{\partial \dot{\eta}}=i\eta^\dagger, &&& \pi_{\eta^\dagger} = \frac{\partial \mathcal{L}}{\partial {\dot{\eta}^\dagger}}=0,
\end{align}
one obtains the following two constraints, 
\begin{align}
  \phi_1 = \pi_\eta-i\eta^\dagger=0, &&&  \phi_2 = \pi_{\eta^\dagger}=0.
\end{align}
The Hamiltonian is given by,
\begin{equation}
  H = i\int dx \, \eta^\dagger \boldsymbol{\sigma}\cdot\boldsymbol{\nabla}\eta+\frac{m}{2}\int dx \left( -\eta^\dagger i\sigma_2\eta^\ast+\eta^Ti\sigma_2\eta \right).
\label{eq:Hamiltonian}
\end{equation}
Following Dirac, 
we aim to obtain the anti-commutation relation, for $\eta$, based on Dirac bracket. 
For instance, the Dirac bracket for $\eta$ and $\eta^\dagger$ is calculated to be,
\begin{multline}
  \{\eta(\bx,t), \eta^\dagger(\by,t)\}_\text{DB}=\{\eta(\bx,t), \eta^\dagger(\by,t)\}_\text{PB}
\\
  -\sum_{a=1,2} \int d^3 x^\prime d^3 y^\prime \{\eta(\bx,t), \phi_a(\bx^\prime ,t)\}_\text{PB} (N^{-1})_{ab}(\bx^\prime, \by^\prime)  \{\phi_b(\by^\prime,t),\eta^\dagger(\by ,t)\}_\text{PB},
\end{multline}
where PB denotes the Poisson bracket and DB the Dirac bracket. 
$N_{ab}$ is a matrix of the Poisson bracket for the constraints,
\begin{equation}
 N_{ab}(\bx, \by) = \{\phi_a(\bx, t), \phi_b(\by,t)\}_\text{PB}= \begin{pmatrix}
    0 & -i \\
    -i & 0
  \end{pmatrix}\delta^{(3)}(\bx-\by) \mathbbm{1}_{2\times 2},
\end{equation}
where $\mathbbm{1}_{2\times 2}$ represents a two by two unit matrix that acts on the two component spinors.
The inverse matrix $(N^{-1})_{ab}$ is calculated to be,
\begin{equation}
  (N^{-1})_{ab} = \begin{pmatrix}
    0 & i \\
    i & 0
  \end{pmatrix}\delta^{(3)}(\bx-\by)\mathbbm{1}_{2\times 2}.
\end{equation}
Then the Dirac brackets among $\eta$ and $\eta^\dagger$ are given as,
\begin{gather}
  \{\eta(\bx,t), \eta^\dagger(\by,t)\}_\text{DB}=-i \delta^{(3)}(\bx-\by) \mathbbm{1}_{2\times 2},
\\
  \{\eta(\bx,t), \eta(\by,t)\}_\text{DB}=\{\eta^\dagger(\bx,t), \eta^\dagger(\by,t)\}_\text{DB}=0.
\end{gather}
Then the anti-commutation relations are,
\begin{gather}
  \{\eta(\bx,t), \eta^\dagger(\by,t)\}=\delta^{(3)}(\bx-\by) \mathbbm{1}_{2\times 2},
 \label{eq:etaAntiCommutationRelation1}
\\
  \{\eta(\bx,t), \eta(\by,t)\}=\{\eta^\dagger(\bx,t), \eta^\dagger(\by,t)\}=0. 
  \label{eq:etaAntiCommutationRelation2}
\end{gather}
Now, we write the two component spinor $\eta$  with the massless, non-zero mode expansion from Eq.(\ref{eq:initialc}) and include a zero mode contribution,
\begin{equation}
  \eta(\bx, t) = \int^\prime \frac{d^3\bp}{(2\pi)^3\sqrt{2\abs{\bp}}}\left(a(\bp, t )\phi_-(\bn_{\bp})e^{i\bp\cdot\bx}-b^\dagger(\bp, t)\phi_-(\bn_{\bp})e^{-i\bp\cdot\bx}\right)+\eta^0(t).
\label{eq:cexpansion}
\end{equation}
In the right-hand side of Eq.(\ref{eq:cexpansion}), the first term represents non-zero modes while the second term represents zero mode defined as,
\begin{equation}
  \int d^3 x  \eta(\bx, t)=V  \eta^{0}(t),
\end{equation}
with $V=(2\pi)^3 \delta^{(3)}(\bp=0) $.
Our goal is to determine the equal-time operator relations among the creation and annihilation operators for both  zero mode and non-zero modes with the knowledge of Eq.(\ref{eq:etaAntiCommutationRelation1}) and Eq.(\ref{eq:etaAntiCommutationRelation2}).
Additionally, we will introduce the zero mode contribution to the lepton number density operator and its expectation value.
The anti-commutation for the zero mode is extracted with Eq.(\ref{eq:etaAntiCommutationRelation1}) by integrating over $\bx$ and $\by$,
\begin{equation}
  \{ \eta^{0}_i(t) , \eta^{0\dagger}_j(t)\}=\frac{1}{V}\delta_{ij},
  \label{eq:zeromodeanticommutation}
\end{equation}
where we denote the two component spinor indices with $i, j=1, 2.$
Next we focus on the non-zero modes.
Recall the meaning of $\int^\prime$, in Eq.(\ref{eq:cexpansion}), is from Eq.(\ref{eq:initialc}) and results in two momenta regions for the spinors.
We use that to explicitly rewrite the integration as a single momenta region, which we label $A$,
\begin{equation}
  \begin{split}
    \eta(\bx,t) =\eta^0(t)+ \int_{\bp\in A} \frac{d^3\bp}{(2\pi)^3\sqrt{2\abs{\bp}}}
    &\left\{[a(\bp, t )\phi_-(\bn_\bp)-b^\dagger(-\bp, t )\phi_-(-\bn_\bp)]e^{i\bp\cdot\bx}\right.
  \\
    &\left.+[a(-\bp, t )\phi_-(-\bn_\bp)-b^\dagger(\bp, t)\phi_-(\bn_\bp)]e^{-i\bp\cdot\bx}\right\}.
  \end{split}
\label{eq:chiral}
\end{equation}
Then, the Fourier transformation is carried out,
\begin{gather}
  \int d^3\bx \, \eta(\bx,t) e^{-i\bq\cdot\bx}=\frac{1}{\sqrt{2|\bq|}}\left[a(\bq, t )\phi_-(\bn_\bq)-b^\dagger(-\bq, t)\phi_-(-\bn_\bq)\right], 
\\
  \int d^3\bx \, \eta(\bx,t) e^{i\bq\cdot\bx}=\frac{1}{\sqrt{2|\bq|}}\left[a(-\bq, t )\phi_-(-\bn_\bq)-b^\dagger(\bq, t)\phi_-(\bn_\bq)\right], 
\end{gather}
where $\bq$ is a momentum in $ A$ region.
This leads to expressions for the operators of non-zero $\bq$, 
\begin{gather}
 \int d^3\bx  \phi^\dagger_-(\bn_\bq)  \eta(\bx, t) e^{-i\bq\cdot\bx} = \frac{1}{\sqrt{2|\bq|}}a(\bq, t ),   \int d^3\bx  \phi^\dagger_-(-\bn_\bq)  \eta(\bx, t) e^{i\bq\cdot\bx} = \frac{1}{\sqrt{2|\bq|}}a(-\bq, t ),
\\
 \int d^3\bx \phi^\dagger_-(\bn_\bq) \eta(\bx, t) e^{i\bq\cdot\bx} = -\frac{1}{\sqrt{2|\bq|}}b^\dagger(\bq, t),  \int d^3\bx \phi^\dagger_-(-\bn_\bq) \eta(\bx, t) e^{-i\bq\cdot\bx} = -\frac{1}{\sqrt{2|\bq|}}b^\dagger(-\bq, t).
\end{gather}
Similar expressions are found for the Hermitian conjugate $\eta^\dagger(t, \bx)$,
\begin{gather}
  \int d^3\bx \, \eta^\dagger(\bx,t)  \phi_-(\bn_\bq) e^{i\bq\cdot\bx} = \frac{1}{\sqrt{2|\bq|}}a^\dagger(\bq, t ),   \int d^3\bx \, \eta^\dagger(\bx,t)  \phi_-(-\bn_\bq) e^{-i\bq\cdot\bx} = \frac{1}{\sqrt{2|\bq|}}a^\dagger(-\bq, t ),
\\
  \int d^3\bx \, \eta^\dagger(\bx,t) \phi_-(\bn_\bq)e^{-i\bq\cdot\bx} =- \frac{1}{\sqrt{2|\bq|}}b(\bq, t),   \int d^3\bx \, \eta^\dagger(\bx,t) \phi_-(-\bn_\bq)e^{i\bq\cdot\bx} =- \frac{1}{\sqrt{2|\bq|}}b(-\bq, t).
\end{gather}
Now we can calculate the equal-time operator relations for $a(\bp, t )$ and $b(\bp, t)$,
\begin{gather}
  \{a(\bq, t ), a^\dagger(\bp, t )\}=(2\pi)^32\abs{\bp}\delta^{(3)}(\bp-\bq),
\\
  \{b(\bq, t ), b^\dagger(\bp, t )\}=(2\pi)^32\abs{\bp}\delta^{(3)}(\bp-\bq),
\end{gather}
with all other relations being zero.
In addition to these anti-commutation relation for non-zero modes, the relation for zero-mode is given in Eq.(\ref{eq:zeromodeanticommutation}).
Notice the non-zero mode relations match Eq.(\ref{eq:masslessAnticommutation}), which were calculated using the Bogoliubov transformation.
We rewrite the Hamiltonian in Eq.(\ref{eq:Hamiltonian}) with the creation and annihilation operators. 
After the normal ordering, the result is
\begin{multline}
  H=H_0+ \int^\prime \frac{d^3 \bp}{(2\pi)^3 2|\bp|}|\bp| [a^\dagger(\bp,t) a(\bp,t)+b^\dagger(\bp, t) b(\bp,t)]
\\
  + m \int_{\bp \in A} \frac{d^3 \bp}{(2\pi)^3 2|\bp|} [-i a(\bp,t)a(-\bp,t)-i b(\bp,t)b(-\bp,t)+\text{h.c.}]
\end{multline}
where $H_0$ is the Hamiltonian for the zero mode and it is defined as,  
\begin{equation}
  H_0= \frac{m}{2} [\eta^{0T}(t) i \sigma_2 \eta^0(t)-\eta^{0\dagger}(t) i \sigma_2 \eta^{0\ast}(t)] V .
\end{equation}
The time evolution of the zero mode reads,
\begin{gather}
  i \frac{d}{dx^0} \eta^0(t) = [\eta^0, H]=m (-i \sigma_2 \eta^{0 \dagger}(t))
  \label{eq:zeromodeevolution}
\\
  i \frac{d}{dx^0} (-i \sigma_2 \eta^{0 \dagger}(t))=m \eta^0(t)
  \label{eq:zeromodeevolutiondagger}
\end{gather}
The solution to both Eq.(\ref{eq:zeromodeevolution}) and Eq.(\ref{eq:zeromodeevolutiondagger}) is given by,
\begin{equation}
  \eta^{0}(t)=\cos m(t-t_0) \eta^{0}(t_0)-i \sin m(t-t_0) (-i \sigma_2 \eta^{0 \dagger}(t_0) ).
  \label{eq:zeromodesol}
\end{equation}
For the non-zero modes, considering $\bp \in A$, the equations of motion are
\begin{gather}
  i \frac{d}{dt}a(\bp, t)=[a(\bp, t), H]=|\bp| a(\bp, t) -i m a^\dagger(-\bp, t),
\\
  i \frac{d}{dt}a^\dagger(-\bp, t)=[a^\dagger(-\bp, t), H]=-|\bp|a^\dagger(-\bp, t) +i m a(\bp, t).
\end{gather}
These lead to solutions consistent with Eq.(\ref{eq:timeDependOpa}) for the multi flavor case,
\begin{gather}
  a(\bp, t)=\left[\cos E(\bp)T -i \frac{|\bp|}{E(\bp)} \sin E(\bp)T \right] a(\bp, t_0)-\frac{m}{E(\bp)} \sin[E(\bp)T] a^\dagger (-\bp, t_0)
  \label{eq:nonzerosolutionplus}
\\
  a^\dagger(-\bp, t)= \left[\cos E(\bp)T +i \frac{|\bp|}{E(\bp)} \sin E(\bp)T\right] a^\dagger (-\bp, t_0)+\frac{m}{E(\bp)} \sin[E(\bp)T] a(\bp, t_0).
  \label{eq:nonzerosolutionminus}
\end{gather}
We used the definition $T\equiv t-t_0$.
The relations consistent with Eq.(\ref{eq:timeDependOpb}) hold for $b(\bp, t)$ and $b^\dagger(-\bp, t)$.
A difference between the zero-mode solution of Eq.(\ref{eq:zeromodesol}) and non-zero-modes of Eq.(\ref{eq:nonzerosolutionplus}) is the amplitude factors.
For the zero-mode cosine and sine amplitudes are not modified by $\abs{\bp}/E(\bp)$ or $m/E(\bp)$.

Next we study the contribution to the lepton number density including the zero-mode.
The zero mode is represented with $\eta^0(t)$  and its
charge  conjugation  $-i \sigma_2\eta^{0 \dagger}(t)$ as shown in Eq.(\ref{eq:chiral}). 
We relate the zero mode to a massive Majorana field.
The Fourier expansion of the massive Majorana field is
\begin{equation}
  \psi_M(\bx, t)= \int \frac{d^3 p}{(2\pi)^3 2 E(\bp)} \sum_{s=\pm1}(a_M(\bp, s) u(\bp,s) e^{-i p\cdot x} + a_M^\dagger(\bp, s) v(\bp,s) e^{i p\cdot x}),
\end{equation}
where $s$ is related to the spin component in an arbitrary direction at the rest frame.
To extract the zero mode from that expansion, we integrate over all space
\begin{equation}
  \psi_M(t)=\frac{\int d^3 x \psi_M(\bx, t)}{V}= \frac{1}{2mV}  \sum_{s=\pm1}(a_M(s) u(s) e^{-i m t } + a_M^\dagger(s) v(s) e^{i m t}) 
\label{eq:zeromodemassiv}
\end{equation}
where the operator is $a_M(s)\equiv a_M(\bp=0,s)$.
We have defined $u(s)\equiv u(\bp=0,s)$ and $v(s)\equiv v(\bp=0,s)$ as the positive and negative energy solutions for the particle at rest.
In spinor/chiral representation of the gamma matrices, they are given as
\begin{align}
  u(s)=\sqrt{m} \begin{pmatrix} \chi^{(s)} \\ \chi^{(s)} \end{pmatrix},
  &&
  v(s)=i \gamma^2 u(s)^\ast=\sqrt{m} \begin{pmatrix} -i \sigma_2\chi^{(s) \ast} \\ i \sigma_2 \chi^{(s) \ast} \end{pmatrix}.
\end{align}
One can choose the two component spinors $\chi^{(s)} (s=\pm 1)$ as spin up or down with respect to an arbitrary direction.
Thus, they satisfy
\begin{equation}
  (\bn \cdot \sigma) \chi^{(s=\pm)}=\pm \chi^{(s=\pm)}.
\label{eq:chi}
\end{equation}
The zero mode of the expansion in Eq.(\ref{eq:zeromodemassiv}) should be matched to the zero mode of Eq.(\ref{eq:psiL}) through the relation
\begin{equation}
  \psi_M(\bx,t)=\nu_L(\bx, t) + (\nu_L(\bx, t))^c.
\end{equation}
That relation is equivalent to Eq.(\ref{eq:MajoranaVSmassless}).
We compare the zero modes of both side of equation to obtain,
\begin{equation}
  \psi_M(t)=\begin{pmatrix} -i \sigma_2 \eta^{0\dagger} (t) \\ \eta^0(t) \end{pmatrix} .
  \label{eq:compa1}
\end{equation}
To facilitate the comparison, we rewrite the right-hand side of Eq.(\ref{eq:compa1}) using Eq.(\ref{eq:zeromodesol}), 
\begin{equation}
  \begin{pmatrix} -i \sigma_2 \eta^{0\dagger} (t) \\ \eta^0(t) \end{pmatrix} = 
  \frac{e^{-i m (t-t_0)}}{2} \begin{pmatrix} \eta^0(t_0)-i \sigma_2 \eta^{0\dagger} (t_0) \\ \eta^0(t_0)-i \sigma_2 \eta^{0\dagger} (t_0) \end{pmatrix}+
  \frac{e^{i m (t-t_0)}}{2} \begin{pmatrix} -(\eta^0(t_0)+i \sigma_2 \eta^{0\dagger} (t_0) ) \\ \eta^0(t_0)+i \sigma_2 \eta^{0\dagger} (t_0) \end{pmatrix}.
\label{eq:RHSofzeromode}
\end{equation}
The left-hand side of Eq.(\ref{eq:compa1}) is given by Eq.(\ref{eq:zeromodemassiv})
\begin{equation}
  \psi_M(t)= \frac{1}{2 \sqrt{m}V}  \sum_{s=\pm1} \left[ a_M(s) 
  \begin{pmatrix} \chi^{(s)} \\ \chi^{(s)} \end{pmatrix} 
  e^{-i m t } + a_M^\dagger(s) 
  \begin{pmatrix} -i \sigma_2\chi^{(s) \ast} \\ i \sigma_2 \chi^{(s) \ast} \end{pmatrix}
  e^{i m t}\right].
\label{eq:LHSofzeromode}
\end{equation}
Comparing both sides of Eq.(\ref{eq:compa1}) given by Eqs.(\ref{eq:RHSofzeromode}) and (\ref{eq:LHSofzeromode}), one obtains,
\begin{eqnarray}
  \eta^0(t_0)&=&\frac{1}{2 \sqrt{m} V}  \sum_{s=\pm 1} \left[ a_M(s) \chi^{(s)} e^{-i m t_0} +a^\dagger_M(s) i \sigma_2 \chi^{(s) \ast} e^{i m t_0} \right],
 \nonumber \\
&=&\frac{1}{2 \sqrt{m} V}  \sum_{s=\pm 1} \left[ a_M(s) \chi^{(s)} e^{-i m t_0} -s a^\dagger_M(s)  \chi^{(-s)} e^{i m t_0} \right].
\label{eq:etazero}
\end{eqnarray}
where we use the relation $ i \sigma_2 \chi^{(s) \ast}=-s \chi^{(-s)} $.
By substituting the two component spinors $\chi^{(s)}$ explicitly, we rewrite Eq.(\ref{eq:etazero})
\begin{equation}
  \eta^0(t_0) = \sum_{s=\pm1} C(s, t_0) \chi^{(s)},
\end{equation}
where the new operator  $C(s,t)$ is given by
\begin{equation}
  C(s,t)=\frac{1}{\sqrt{2m} V} \frac{a_M(s) e^{-i m t }+sa_M^{\dagger}(-s) e^{i m t}}{\sqrt{2}}.
\end{equation}
That obeys the anti-commutation relation following Eq.(\ref{eq:zeromodeanticommutation}),
\begin{equation}
\{C(s,t), C^\dagger(s',t)\}=\frac{1}{V}\delta_{ss'}.
\end{equation}
Additionally, it satisfies the differential equation,
\begin{equation}
  i\frac{dC(s,t)}{dt}=-smC^\dagger(-s,t).
\end{equation}
The solution is given by the operator defined at $t=t_0$ as,
\begin{equation}
  C(s,t)=\cos m(t-t_0) C(s,t_0)+ i s \sin m(t-t_0) C^\dagger(-s,t_0)
  \label{eq:coperatortimeevo}
\end{equation}
This completes the expansion of Eq.(\ref{eq:chiral}) in terms of the operators for the non-zero modes $a(\pm\bp,t)$ and $b(\pm\bp,t)$, and the zero mode operator $C(s,t)$.
The result is,
\begin{equation}
  \begin{split}
    \eta(\bx,t) =\sum_{s=\pm1}C(s,t)\chi^{(s)} + \int_{\bp\in A} \frac{d^3\bp}{(2\pi)^3\sqrt{2\abs{\bp}}}
    &\left\{[a(\bp, t )\phi_-(\bn_\bp)-b^\dagger(-\bp, t )\phi_-(-\bn_\bp)]e^{i\bp\cdot\bx}\right.
  \\
    &\left.+[a(-\bp, t )\phi_-(-\bn_\bp)-b^\dagger(\bp, t)\phi_-(\bn_\bp)]e^{-i\bp\cdot\bx}\right\}.
  \end{split}
\end{equation}

Next, we consider the lepton number density operator including the zero mode contribution.
This can be view as an extension of Eq.(\ref{Eq:MajoranaDensity}) for a single flavor,
\begin{multline}
  l^M(t,\bx)= :\overline{\nu_L}(\bx, t) \gamma^0 {\nu_L}(\bx,t): =\sum_{s=\pm1}C^\dagger(s,t) C(s,t)
\\
  +\int^\prime \frac{d^3 p}{(2\pi)^3 \sqrt{2|\bp|}} \sum_{s=\pm1} C^\dagger(s,t) (a(\bp, t) e^{i \bp \cdot \bx}- b^\dagger(\bp, t) e^{-i \bp \cdot \bx}) (\chi^{(s)\dagger}  \cdot \phi_-(n_\bp))
\\
  +\int^\prime \frac{d^3 k}{(2\pi)^3 \sqrt{2|\bk|}} \sum_{s=\pm1} ( \phi(n_\bk)^\dagger \cdot \chi^{(s)} )(a^\dagger(\bk, t) e^{-i \bk \cdot \bx}- b(\bk, t) e^{i \bk \cdot \bx})  C(s,t)
\\
  +\int^\prime \frac{d^3 k}{(2\pi)^3 \sqrt{2|\bk|}} \int^\prime \frac{d^3 p}{(2\pi)^3 \sqrt{2|\bp|}} ( \phi_-(n_\bk)^\dagger  \cdot \phi_-(n_\bp)) 
\\
  \times:(a^\dagger(\bk, t) e^{-i \bk \cdot \bx}- b(\bk, t) e^{i \bk \cdot \bx}) (a(\bp, t) e^{i \bp \cdot \bx}- b^\dagger(\bp, t) e^{-i \bp \cdot \bx}):
 \label{eq:LNDWithZERO}
\end{multline}
The first term of Eq.(\ref{eq:LNDWithZERO}) comes from the zero mode.
It is given by the operator defined at $t=t_0$ in Eq.(\ref{eq:coperatortimeevo}) to be,
\begin{multline}
\sum_{s=\pm1}C^\dagger(s,t) C(s,t)=\frac{1}{V}(1-\cos2m(t-t_0))+ \cos2m(t-t_0) \sum_{s=\pm1} C^\dagger(s,t_0) C(s,t_0)
\\
-i  \{C(-1,t_0) C(1,t_0)-C^\dagger(1,t_0) C^\dagger(-1,t_0) \} 
\sin2m(t-t_0). 
\end{multline}
The second and third lines of Eq.(\ref{eq:LNDWithZERO}) are more complicated, because they are a mixture of zero with non-zero modes.
Line four of Eq.(\ref{eq:LNDWithZERO}) is equivalent to a single flavor version of Eq.(\ref{Eq:MajoranaDensity}).
We take the expectation value of the lepton number density with the following momentum distribution
\begin{equation}
  \vert \Psi(q^0, \sigma_q, c^0(s_0)), t_0 \rangle = \vert \psi(q^0;\sigma_q), t_0 \rangle +  c^0(s_0) C^\dagger(s_0, t_0) \vert 0(t_0) \rangle
  \label{eq:densitystatewithzeromode}
\end{equation}
where $\vert \psi(q^0;\sigma_q), t_0 \rangle$ is similar to the initial Gaussian distribution of Eq.(\ref{Eq:Wavepacket}).
It is given by a superposition of the non-zero modes,
\begin{equation}
  \lvert \psi(q^0;\sigma_q), t_0 \rangle= \frac{1}{\sqrt{\sigma_q}(2\pi)^{3/4}}\int'\frac{dq}{\sqrt{A} \sqrt{2|q|}}e^{-\frac{(q-q^0)^2}{4\sigma_q^2}} a^\dagger(\bq, t_0) \lvert 0(t_0) \rangle ;
  \label{eq:singleflavorgaussian}
\end{equation}
where $\bq=(0,q,0)$.
In addition to non-zero mode contribution, we add the zero mode contribution with the amplitude $c^0(s)$.
Note that the dimension of the coefficient is $\sqrt{V}$.
The expectation value of the  lepton number densities given by the sum of the following terms
\begin{multline}
  \langle \Psi(q^0, \sigma_q, c^0(s_0)), t_0 \rvert l^M(t,\bx) \lvert \Psi(q^0, \sigma_q, c^0(s_0)), t_0  \rangle =\langle  \psi(q^0;\sigma_q), t_0 \rvert l^M(t,\bx) \lvert  \psi(q^0;\sigma_q), t_0 \rangle
\\
  \begin{aligned}
    &+c^0(s_0)\langle  \psi(q^0;\sigma_q), t_0 \rvert l^M(t,\bx)   C^\dagger(s_0, t_0) \lvert 0(t_0) \rangle  + \text{h.c.}
  \\
    &+ |c^0(s_0)|^2  \langle 0(t_0)  \rvert  C(s_0, t_0)   l^M(t,\bx)   C^\dagger(s_0, t_0) \lvert 0(t_0) \rangle.
  \end{aligned}
  \label{eq:zeromodeexpectationvalue}
\end{multline}
We start with the first line by substituting Eq.(\ref{eq:LNDWithZERO}),
\begin{multline}
  \langle  \psi(q^0;\sigma_q), t_0 \rvert l^M(t,\bx) \lvert  \psi(q^0;\sigma_q), t_0 \rangle =\frac{1-\cos2m(t-t_0)}{V}  \langle  \psi(q^0;\sigma_q), t_0 \rvert \psi(q^0;\sigma_q), t_0 \rangle
\\
  + \frac{1}{\sigma_q(2\pi)^{3/2}} \iint' \frac{dq'dq}{A} e^{-\frac{(q'-q^0)^2+(q-q^0)^2}{4\sigma_q^2}-i(q'-q) x^2 }  \left[-\frac{m}{E(q')}\sin{E(q')T}\frac{m}{E(q)}\sin{E(q)T} \right.
\\
  \left. + \left(\cos{E(q')T}+i\frac{|q'|}{E(q')}\sin{E(q')T}\right) \left(\cos{E(q)T}-i\frac{|q|}{E(q)}\sin{E(q)T}\right) \right].
\end{multline}
Although, this result is similar to main text of Eq.(\ref{Eq:kpExpectationValue}), the term proportional to $1-\cos2m(t-t_0)$ is from the zero mode.
For the last line of Eq.(\ref{eq:zeromodeexpectationvalue}), we are able to compute the matrix elements with
\begin{gather}
  \langle 0(t_0) \vert C(s_0,t_0) \sum_{s=\pm1}C^\dagger(s,t) C(s,t) C^\dagger(s_0,t_0) \vert 0(t_0) \rangle =\frac{1}{V^2}.
\\
 |c^0(s_0)|^2  \langle 0(t_0)  \rvert  C(s_0, t_0)   l^M(t,\bx)   C^\dagger(s_0, t_0) \lvert 0(t_0) \rangle=  \frac{|c^0(s_0)|^2}{V^2}.
\end{gather}
Finally, the third line of  Eq.(\ref{eq:LNDWithZERO}) is computed to be,
\begin{equation}
  \begin{split}
    &\langle  \psi(q^0;\sigma_q), t_0 \rvert  \int^\prime \frac{d^3 k}{(2\pi)^3 \sqrt{2|\bk|}} \sum_{s=\pm1} ( \phi_-(n_\bk)^\dagger \cdot \chi^{(s)} )  (a^\dagger(\bk, t) e^{-i \bk \cdot \bx}- b(\bk, t) e^{i \bk \cdot \bx})  C(s,t) C^\dagger(s_0, t_0) \vert 0(t_0) \rangle,
  \\
    &= \int^\prime \frac{d^3 k}{(2\pi)^3 \sqrt{2|\bk|}}  \sum_{s=\pm1}( \phi_-(n_\bk)^\dagger \cdot \chi^{(s)} ) \langle \psi(q^0, \sigma), t_0 \rvert  a^\dagger(\bk, t)  \vert 0(t_0) \rangle e^{-i \bk \cdot \bx}\langle 0(t_0) \vert C(s,t) C^\dagger(s_0,t_0) \vert 0(t_0) \rangle,
  \\
    &=\int^\prime \frac{d^3 k}{(2\pi)^3 \sqrt{2|\bk|}} ( \phi_-(n_\bk)^\dagger \cdot \chi^{(s_0)} )  \langle  \psi(q^0;\sigma_q), t_0 \rvert   a^\dagger(\bk, t)  \vert 0(t_0) \rangle e^{-i \bk \cdot \bx} \frac{1}{V} \cos m(t-t_0).
  \end{split}
\end{equation}
We perform the integral over $\bk$ using the definition of Eq.(\ref{eq:singleflavorgaussian}) for the bra- vector,
\begin{equation}
  \begin{split}
    \int^\prime &\frac{d^3 k}{(2\pi)^3 \sqrt{2|\bk|}} ( \phi_-(n_\bk)^\dagger \cdot \chi^{(s_0)} )  \langle  \psi(q^0;\sigma_q), t_0 \rvert   a^\dagger(\bk, t)  \vert 0(t_0) \rangle e^{-i \bk \cdot \bx}
\\
    &=\int^\prime \frac{d^3 k}{(2\pi)^3 \sqrt{2|\bk|}}( \phi_-(n_\bk)^\dagger \cdot \chi^{(s_0)} ) \int'\frac{dq e^{-\frac{(q-q^0)^2}{4\sigma_q^2} -i \bk \cdot \bx }}{\sqrt{A \sigma_q (2\pi)^{3/2}} \sqrt{2|q|}} \langle 0(t_0) \vert  a(\bq, t_0) a^\dagger(\bk, t)  \vert 0(t_0) \rangle
\\
    &= \int_{\bk \in A} \frac{d^3 k}{(2\pi)^3 \sqrt{2|\bk|}}( \phi_-(n_\bk)^\dagger \cdot \chi^{(s_0)} ) \int^{+\infty}_{+0}\frac{dq e^{-\frac{(q-q^0)^2}{4\sigma_q^2}-i q \bf{e}_2 \cdot \bx} (\cos E(q) t+ i\frac{q}{E(q)} \sin E(q) t)}{\sqrt{A \sigma_q (2\pi)^{3/2}} \sqrt{2|q|}}
\\
    &\times (2\pi)^3 2|q| \delta(k^1) \delta(k^2-q) \delta(k^3)
\\
    &+\int_{\bk \in \overline{A}} \frac{d^3 k}{(2\pi)^3 \sqrt{2|\bk|}} ( \phi_-(n_\bk)^\dagger \cdot \chi^{(s_0)} )\int^{-0}_{-\infty}\frac{dq e^{-\frac{(q-q^0)^2}{4\sigma_q^2}-i q \bf{e}_2 \cdot \bx}  (\cos E(q) t+ i\frac{|q|}{E(q)} \sin E(q) t)  }{\sqrt{A \sigma_q (2\pi)^{3/2}} \sqrt{2|q|}}
\\
    &\times(2\pi)^3 2|q| \delta(k^1) \delta(k^2-q) \delta(k^3)
\\
    &=(\phi_-({\bf{e}}_2)^\dagger \cdot \chi^{(s_0)} ) \int^{+\infty}_{+0} \frac{dq}{\sqrt{A \sigma_q (2\pi)^{3/2} }}e^{-\frac{(q-q^0)^2}{4\sigma_q^2}-i q \bf{e}_2 \cdot \bx  } (\cos E(q) t+ i\frac{|q|}{E(q)} \sin E(q) t)
\\
    &+( \phi_-(-{\bf{e}}_2)^\dagger \cdot \chi^{(s_0)} )  \int^{-0}_{-\infty}  \frac{dq}{\sqrt{A \sigma_q (2\pi)^{3/2} }}e^{-\frac{(q-q^0)^2}{4\sigma_q^2}-i q \bf{e}_2 \cdot \bx} (\cos E(q) t+ i\frac{|q|}{E(q)} \sin E(q) t),
  \end{split}
\end{equation}
where $\int^{+\infty}_{+0} dq $ implies integration over positive $q$ and $ \int^{-0}_{-\infty} dq $ for negative $q$.
If we choose $\chi^{(s_0)}$ such that $\bn=\bf{e}_2$ in Eq.(\ref{eq:chi}),
\begin{align}
  \phi_-({\bf{e}}_2)^\dagger \cdot \chi^{(s_0=1)}=0 , && \phi_-({\bf{e}}_2)^\dagger \cdot \chi^{(s_0=-1)}=1
\\
  \phi_-(-{\bf{e}}_2)^\dagger \cdot \chi^{(s_0=1)}=-i, && \phi_-(-{\bf{e}}_2)^\dagger \cdot \chi^{(s_0=-1)}=0.
\end{align}
By adding all the contribution,  the expectation value for the lepton number density Eq.(\ref{eq:LNDWithZERO}) given by the matrix element with the state Eq.(\ref{eq:densitystatewithzeromode}) is
\begin{multline}
  \langle \Psi(q^0, \sigma_q, c^0(s_0=1)) \vert l^M(t,\bx) \vert \Psi(q^0, \sigma_q, c^0(s_0=1)) \rangle = \frac{1-\cos2m(t-t_0)}{V}  +\frac{|c^0(s_0=1)|^2}{V^2}
\\
  -i \frac{c^0(s_0=1) \cos m T }{V} \int^{-0}_{-\infty} \frac{dq}{\sqrt{A \sigma_q (2\pi)^{3/2} 2}}(\cos( E(q)T )+i \frac{q}{E(q)} \sin( E(q)T )) 
 (e^{-\frac{(q-q^0)^2}{4\sigma_q^2}-i q x^2} )
\\
  + i \frac{{c^{0}(s_0=1)}^\ast \cos m T }{V} \int^{-0}_{-\infty} \frac{dq}{\sqrt{A \sigma_q (2\pi)^{3/2} 2}}(\cos( E(q)T)-i \frac{q}{E(q)} \sin( E(q)T )) (e^{-\frac{(q-q^0)^2}{4\sigma_q^2}+i q x^2})
\\
  +\frac{1}{\sigma_q(2\pi)^{3/2}} \iint' \frac{dq'dq}{A}e^{-\frac{(q'-q^0)^2+(q-q^0)^2}{4\sigma_q^2}-i(q'-q) x^2 }
\\
  \times\left[\left(\cos{E(q')T}+i\frac{|q'|}{E(q')}\sin{E(q')T}\right) \left(\cos{E(q)T}-i\frac{|q|}{E(q)}\sin{E(q)T}\right) \right.
\\
  \left. -\frac{m}{E(q')}\sin{E(q')T}\frac{m}{E(q)}\sin{E(q)T}\right];
\label{Eq:ExpectationValueWZERO}
\end{multline}
For the opposite direction of the spin state of the zero mode $s_0=-1$, the result is,
\begin{multline}
  \langle \Psi(q^0, \sigma_q, c^0(s_0=-1)) \vert l^M(t,\bx) \vert \Psi(q^0, \sigma_q, c^0(s_0=-1)) \rangle =\frac{1-\cos2m(t-t_0)}{V}  +\frac{|c^0(s_0=-1)|^2}{V^2}
\\
  +\frac{c^0(s_0=-1) \cos m T }{V} \int_{+0}^{+\infty} \frac{dq}{\sqrt{A \sigma_q (2\pi)^{3/2} 2}} ( \cos E(q)T +i \frac{q}{E(q)} \sin E(q) T)  (e^{-\frac{(q-q^0)^2}{4\sigma_q^2}-i q x^2} )
\\
  + \frac{{c^{0}(s_0=-1)}^\ast \cos m T }{V} \int_{+0}^{+\infty} \frac{dq}{\sqrt{A \sigma_q (2\pi)^{3/2} 2}} ( \cos E(q)T -i \frac{q}{E(q)} \sin E(q) T)(e^{-\frac{(q-q^0)^2}{4\sigma_q^2}+i q x^2 })
\\
  +\frac{1}{\sigma_q(2\pi)^{3/2}} \iint' \frac{dq'dq}{A} e^{-\frac{(q'-q^0)^2+(q-q^0)^2}{4\sigma_q^2}-i(q'-q) x^2 }
\\
  \times\left[\left(\cos{E(q')T}+i\frac{|q'|}{E(q')}\sin{E(q')T}\right) \left(\cos{E(q)T}-i\frac{|q|}{E(q)}\sin{E(q)T}\right) \right.
\\
  \left. -\frac{m}{E(q')}\sin{E(q')T}\frac{m}{E(q)}\sin{E(q)T}\right];
\label{Eq:ExpectationValueWZEROs0m}
\end{multline}
Both Eq.(\ref{Eq:ExpectationValueWZERO}) and Eq.(\ref{Eq:ExpectationValueWZEROs0m}) imply that even if the zero mode amplitude $c^0(s_0=\pm1)$ vanishes in the initial state, the expectation value has the new contribution.
To examine the effect, we compute the linear density as done in the text by integrating the lepton number density over the area $A=\int \int dx^1 dx^3$.
The result for Eq.(\ref{Eq:ExpectationValueWZEROs0m}) is,
\begin{multline}
\lambda^M(T=t-t_0, x^2)=
\int dx^1 dx^3  \langle \Psi(q^0, \sigma_q, c^0(s_0=-1)) \vert l^M(t,\bx) \vert \Psi(q^0, \sigma_q, c^0(s_0=-1)) \rangle \\
=\frac{1-\cos2m(t-t_0)}{L}  +\frac{|c^0(s_0=-1)|^2}{V L}
\\
  +\frac{c^0(s_0=-1) \cos m T }{L\sqrt{A}} \int_{0+}^\infty \frac{dq}{\sqrt{ \sigma_q (2\pi)^{3/2} 2}} ( \cos E(q)T +i \frac{q}{E(q)} \sin E(q) T)  (e^{-\frac{(q-q^0)^2}{4\sigma_q^2}-i q x^2} )
\\
  + \frac{{c^{0}(s_0=-1)}^\ast \cos m T }{L \sqrt{A}} \int_{0+}^\infty \frac{dq}{\sqrt{ \sigma_q (2\pi)^{3/2} 2}} ( \cos E(q)T -i \frac{q}{E(q)} \sin E(q) T)(e^{-\frac{(q-q^0)^2}{4\sigma_q^2}+i q x^2 })
\\
  +\frac{1}{\sigma_q(2\pi)^{3/2}} \iint' {dq'dq} e^{-\frac{(q'-q^0)^2+(q-q^0)^2}{4\sigma_q^2}-i(q'-q) x^2 }
\\
  \times\left[\left(\cos{E(q')T}+i\frac{|q'|}{E(q')}\sin{E(q')T}\right) \left(\cos{E(q)T}-i\frac{|q|}{E(q)}\sin{E(q)T}\right) \right.
\\
  \left. -\frac{m}{E(q')}\sin{E(q')T}\frac{m}{E(q)}\sin{E(q)T}\right];
\end{multline}
where we use $V=AL$. 


\end{document}